\documentclass[12pt]{article}
\usepackage{epsfig} \usepackage{amssymb} \usepackage{amsfonts}
\usepackage{psfig}

\newcommand{\zz}{&&\hspace*{-4cm}}
\newcommand{\ep}{\varepsilon}

\newcommand{\titul}[1] {\begin{center}{\large\bf #1 }
\end{center}\vskip 1.cm}

\newcommand{\abstr}[1] {{\begin{center} \vskip .5cm {\bf Abstract
                        \vspace{0pt}} \end{center}}\begin{quote} #1
                        \end{quote}}
\newcommand{\bea}{\begin{eqnarray}}
\newcommand{\eea}{\end{eqnarray}}
\newcommand{\MSbar}{\overline{\rm MS}}
\newcommand{\as}{\alpha_s}
\newcommand{\asMZ}{\alpha_s(M^2_Z)}
\newcommand{\ar}{\overline a_s}

\begin{document}

\begin{titlepage}
%\prepr{
%%DESY 01-xxx\\
%JINR E2-01-xxx\\
%27 August 2001\\Draft 4.20}
%
\vskip 1cm
\titul{
A systematic study of QCD coupling constant from \\deep inelastic
measurements}

\begin{center}
{\bf V.G. Krivokhijine
%and A.V. Kotikov 
}\\ [0.5cm]

{\em Laboratory of Particle Physics, Joint Institute for Nuclear
Research,\\ 141980 Dubna, Russia}\\

\vskip 0.5cm

and 

\vskip 0.5cm

{\bf A.V. Kotikov 
}\\ [0.5cm]

{\em Bogoliubov Laboratory of Theoretical Physics, Joint Institute for Nuclear
Research,\\ 141980 Dubna, Russia}\\

\end{center}

\abstr{
We reanalyze 
%high statistic 
deep inelastic scattering data 
of BCDMS Collaboration by including proper cuts of  ranges
with large systematic errors. 
We perform also the 
%combine 
fits of high statistic deep inelastic scattering data 
of BCDMS, SLAC, NM and BFP Collaborations  
taking the data separately and in combined way and find good agreement
between these analyses. We
%and 
extract the values of both
the QCD coupling constant $\alpha_s(M^2_Z)$ up to
NLO level and of the power corrections to the 
%proton 
structure function $F_2$.
The fits of the combined data for
the nonsinglet part of the structure function $F_2$
predict the coupling constant value $\alpha_s(M^2_Z) = 0.1174 \pm 0.0007$
(stat) $\pm 0.0019$ (syst) $\pm 0.0010$ (normalization) 
(or QCD parameter 
$\Lambda^{(5)}_{\overline{MS}} = (204
\pm 25~\mbox{(total exper.err.)}) MeV$).
The fits of the combined data for
both: the nonsinglet part and the singlet one, lead to the
%little higher 
values $\alpha_s(M^2_Z) = 0.1177 \pm 0.0007$
(stat) $\pm 0.0021$ (syst) $\pm 0.0009$ (normalization) 
(or QCD parameter 
$\Lambda^{(5)}_{\overline{MS}} = (208
\pm 27~\mbox{(total exper.err.)}) MeV$).
Both above values are  in very good
 agreement with each
%one to 
other.
% even (? if statistical errors are included only ?). 
We estimate theoretical uncertainties for $\alpha_s(M^2_Z)$ as $+ 0.0047$ and 
$- 0.0057$ from fits of the combine data, 
when complete singlet and nonsinglet $Q^2$ evolution is
taken into account.\\

$PACS:~~12.38~Aw,\,Bx,\,Qk$\\

}

\end{titlepage}

\section{ Introduction }

The deep inelastic scattering (DIS) leptons on hadrons is the basical
 process to study the values of the parton distribution functions
which are universal (after choosing of factorization and renormalization 
schemes) and
can be used in other processes.
The accuracy of the present data for deep inelastic
%(DIS) 
structure functions (SF) reached the level at which
the $Q^2$-dependence of logarithmic QCD-motivated and power-like ones
%are observed and 
may be studied separately 
%(see, for example, the recent reviews 
(for a review, see the recent papers \cite{Maul, Beneke} and references 
therein).

In the present article we analyze at the next-to-leading (NLO)
 order \footnote{The evaluation
of $\alpha_s^3(Q^2)$ corrections to anomalous dimensions of Wilson operators,
that will be done in nearest future by Vermaseren and his coauthors (see 
discussions in \cite{Stirling}), 
gives a possibility
to apply many modern programs to perform fits of data at next-next-to-leading
order (NNLO) of perturbative theory (see detail discussions in 
%Section 5
Summary).}
%large experience in fits of data
of perturbative QCD
the most known DIS SF 
$F_2(x,Q^2)$ taking into account SLAC, NMC,  BCDMS
and BFP
experimental data \cite{SLAC1}-\cite{BFP}.
We
stress the power-like effects, so-called twist-4 (i.e.
$\sim 1/Q^2$) 
%and twist-6 (i.e. $\sim 1/Q^4$) 
contributions.
To our purposes we represent the SF $F_2(x,Q^2)$ as the contribution
of the leading twist part $F_2^{pQCD}(x,Q^2)$ 
described by perturbative QCD and the  
nonperturbative part (twist-four terms $\sim 1/Q^2$):
\begin{equation}
F_2(x,Q^2) 
\equiv F_2^{full}(x,Q^2)
=F_2^{pQCD}(x,Q^2)+\frac{\tilde h_4(x)}{Q^2} 
%+ \frac{\tilde h_6(x,Q^2)}{Q^4}
\label{1}
\end{equation}

The SF $F_2^{pQCD}(x,Q^2)$ obeys the (leading twist)
perturbative QCD dynamics
including the target mass corrections (TMC)
(and coincides with $F_2^{tw2}(x,Q^2)$
when the target mass corrections are withdrawn).

The Eq.(\ref{1}) allows us to separate pure kinematical power corrections,
i.e. TMC, so that the function $\tilde h_4(x)$ corresponds to ``dynamical''
contribution of the twist-four operators. The parameterization (\ref{1})
implies 
\footnote{The r.h.s. of the Eq.(\ref{1}) is represented sometimes
as  $F_2^{pQCD}(x,Q^2)(1+\overline h_4(x)/Q^2)$. It implies that the
anomalous dimensions of the twist-two and twist-four operators are equal
to each other.} that the anomalous dimensions of the twist-four operators 
are equal to zero, that is not correct in principle. Moreover, there
are estimations of these anomalous dimensions (see \cite{BuKuLi}). Meanwhile,
in view of limited precision of the data, the approximation (\ref{1})
and one in the footnote 2 give rather good predictions (see discussions in 
\cite{Al2000,Al2001}).\\

Contrary to standard fits (see, for example, \cite{ViMi,fits}) when the direct
numerical calculations based on Dokshitzer-Gribov-Lipatov-Altarelli-Parisi
(DGLAP) equation \cite{DGLAP} are used to evaluate structure functions, 
we use the exact solution of DGLAP equation
for the Mellin moments $M_n^{tw2}(Q^2)$ of
%$F_2^{full}(x,Q^2)$, $F_2^{pQCD}(x,Q^2)$ and
SF $F_2^{tw2}(x,Q^2)$:
\begin{equation}
M_n^{k}(Q^2)=\int_0^1 x^{n-2}\,F_2^{k}(x,Q^2)\,dx~~~~~~~ (\mbox{hereafter }
k=full, pQCD, tw2, ...)
\label{2}
\end{equation}
and
the subsequent reproduction of $F_2^{full}(x,Q^2)$, $F_2^{pQCD}(x,Q^2)$  and/or
$F_2^{tw2}(x,Q^2)$ at every needed $Q^2$-value with help of the Jacobi 
Polynomial expansion method \cite{Barker}-\cite{Kri1}
(see similar analyses at the NLO level 
\cite{Kri}-\cite{KKDIS}
and at the NNLO level and above \cite{PKK}-\cite{KPS1}).

%{\bf 2.}~
The method of the Jacobi polynomial expansion was developed
in \cite{Barker, Kri} and described in details
in Refs.\cite{Kri1}. Here we consider only
some basical definitions in Section 3.\\

The paper has the following structure: in Section 2 we present basic formulae,
which are needed in our analyses: we consider different types of 
$Q^2$-dependence of SF moments, effects of nuclear corrections and 
heavy quark thresholds,
the structure of normalization of parton densities in singlet and nonsinglet
channels. In Section 3 we introduce the basic elements of our fits. Sections
4 and 5 contain conditions and results of several types of fits with
the nonsinglet and singlet evolutions
%fits of 
for different sets of data. 
In Section 6 we study the dependence of the results on choice of factorization
and renormalization scales.
In Section 7 we summarize the basic 
observations 
following from the fits and discuss possible future extensions of the
analyses.

\section{ $Q^2$ dependence of SF and their moments }

%First of all, 
In this section we analyze Eq.(\ref{1}) in detail, considering separately
different types of $Q^2$-dependence of structure function $F_2$.

\subsection{ The leading-twist $Q^2$ dependence }

To study the $Q^2$-dependence of the SF $F_2^{tw2}(x,Q^2)= F_2^{NS}(x,Q^2)
+ F_2^{S}(x,Q^2)$, which splits explicitly into the nonsinglet (NS) part and 
the singlet (S) one, it is very useful to introduce parton
distribution functions (PDF)\footnote{Our PDF are multiplied by $x$ to compare
with standard definition.}:
gluon one $f_G(x,Q^2)$ and singlet and nonsinglet quarks ones
$f_S(x,Q^2)$ and $f_{NS}(x,Q^2)$.

The moments $M_n^{NS}(Q^2)$ and $M_n^{S}(Q^2)$ 
of NS and S parts of SF $F_2$ (see Eq.(\ref{2}) for definition) are
connected with the corresponding
moments of PDF $f_i(x,Q^2)$ (hereafter $i=NS,S,G$)
$$
f_i(n,Q^2)=\int_0^1 x^{n-2}f_i(n,Q^2)dx
$$
in the following way (see \cite{Buras}, for example) 
\begin{eqnarray}
M_n^{NS}(Q^2) &=& K_{NS}(f) \cdot
C_{NS}^{tw2}(n, \ar(Q^2)) \cdot f_{NS}(n,Q^2) \nonumber \\
%\label{3.a} \\
& &  \label{3.a} \\
M_n^{S}(Q^2) &=& K_{S}(f) \cdot \biggl[
C_{S}^{tw2}(n, \ar(Q^2)) \cdot f_{S}(n,Q^2)
+ C_{G}^{tw2}(n, \ar(Q^2)) \cdot f_{G}(n,Q^2) \biggr], 
\nonumber 
%\label{3.b}
\end{eqnarray}
where\footnote{Sometimes the last term
$K_{S}(f) \cdot C_{G}^{tw2}(n, \ar(Q^2)) \cdot f_{G}(n,Q^2)$
we will call as gluon part of singlet moment and denote it as 
$M_n^{G}(Q^2)$.}
%$\ar(Q^2) = \alpha_s(Q^2)/(4\pi)$ 
\bea
\ar(Q^2) ~=~ \frac{\alpha_s(Q^2)}{4\pi}
\label{coupl}
\eea
and $C_{i}^{tw2}(n, \ar(Q^2))$
are so-called Wilson coefficient functions.
We have introduced here also the coefficients
\bea
K_{S}(f) ~=~ \sum^{f}_{m=1} e^2_m /f, ~~~  ~~~
K_{NS}(f) ~=~ e^2_u - K_{S}(f),
%\sum^{f}_{m=1} e^2_m /f
\label{3.c}
\eea
which come from definition of SF $F_2$ (see, for example, \cite{Buras}).
%where 
Here $f$ is the number of  active quarks and
$e^2_{m}$ is charge square of the active quark of $m$ flavor.\\ 
%when there are active $f$ quarks. \\

{\bf 2.1.1}~~ 
The dependence of $\ar(Q^2)$ is given by the renormalization group
equation, which in the NLO QCD approximation reads:
\bea
\frac{1}{\ar(Q^2)} - \frac{1}{\ar(M_Z^2)} + 
\frac{\beta_1}{\beta_0} 
\ln{\left[\frac{\ar(Q^2)}{\ar(M_Z^2)}
\frac{\left(\beta_0+ \beta_1 \ar(M_Z^2)\right)}
{\left(\beta_0+ \beta_1 \ar(Q^2)\right)}\right]} = 
\beta_0 
\ln{\left(\frac{Q^2}{M_Z^2}\right)}
\label{1.co},
\eea
where $\ar(M_Z^2)$ is served as a normalization. Here and below we use 
$\beta_0$ and $\beta_1$
for the first and the second terms 
with respect to $\ar $ of QCD $\beta$-function:
\bea
\beta(\ar) ~=~ -\beta_0 \ar^2 - \beta_1 \ar^3 ~+~ ...~~ .
\nonumber 
\eea

The equation (\ref{1.co}) allows us to eliminate the QCD parameter 
$\Lambda_{QCD}$ from our analysis. However, sometimes we will present it
in our discussions, essentially to compare it with the results of old
fits. The coupling constant $\ar(Q^2)$ is expressed through $\Lambda_{QCD}$
(in $\MSbar$ scheme, where $\Lambda_{QCD}=\Lambda_{\MSbar}$) as
\bea
\frac{1}{\ar(Q^2)} + 
\frac{\beta_1}{\beta_0} 
\ln{\left[\frac{\beta_0^2 \ar(Q^2)}
{\left(\beta_0+ \beta_1 \ar(Q^2)\right)}\right]} = 
\beta_0 
\ln{\left(\frac{Q^2}{\Lambda^2_{\MSbar}}\right)}
\label{2.co},
\eea

The relation between the normalization $\ar(M_Z^2)$ and the QCD parameter 
$\Lambda_{QCD}$ can be obtained from Eq.(\ref{2.co}) with the replacement
$Q^2 \to M_Z^2$.

We would like to note that the approximations of Eq.(\ref{2.co}), based on
the expansion of inverse powers of 
$\ln{\left(Q^2/\Lambda^2_{\MSbar}\right)}$
are very popular. The accuracy of these expansions for evolution of $\ar$
from $O$(GeV$^2$) to $M_Z^2$ may be as large as $0.001$ \cite{Al2001},
which is comparable with the experimental uncertainties of the $\as(M_Z^2)$
value extracted from the data (see our analyses in Sections 4 and 5). 

Note also that sometimes (see, for example, \cite{PKK}) the equation
\bea
\frac{1}{\ar(Q^2)} + 
\frac{\beta_1}{\beta_0} 
\ln{\left(\beta_0 \ar(Q^2) \right)} = 
\beta_0 
\ln{\left(\frac{Q^2}{\Lambda^2_{\MSbar}}\right)}
\label{3.co},
\eea
is used in analyses. This equation can be obtained from the
basic equation
\bea
\ln{\left(\frac{Q^2}{\Lambda^2_{\MSbar}}\right)}~=~ 
\int^{\ar(Q^2)}  
\frac{db}{\beta(b)}, 
\label{4.co}
\eea
by expansion of inverse QCD $\beta$-function in r.h.s. of (\ref{4.co})
$1/\beta(\ar)$ in powers of $\ar$. The difference between Eqs.(\ref{3.co})
and (\ref{2.co}) may be as large as $0.001$ at $O$(GeV$^2$) range.
In order to escape the above uncertainties we
use in the analyses the exact numerical solution (with accuracy about 
$10^{-5}$) of Eq.(\ref{1.co}) instead. For recalculation of the QCD
parameter $\Lambda_{\MSbar}$ from $\Lambda^{(f)}_{\MSbar}$ to
$\Lambda^{(f\pm1)}_{\MSbar}$ (i.e. from $\Lambda_{\MSbar}$ at $f$ active
quark flavors to $\Lambda_{\MSbar}$ at $f\pm 1$ active
quark flavors), because $\beta_0$ and $\beta_1$ are $f$-dependent functions,
we use formulae at NLO approximation from 
Ref. \cite{Chetyrkin} (see discussions in the subsection 2.4).\\

{\bf 2.1.2}~~
The coefficient functions $C_{i}^{tw2}(n, \ar(Q^2))$ ($i=NS,S,G$)
have the
following form
\bea
C_{i}^{tw2}(n, \ar(Q^2)) = 1 - \delta^G_i + \ar \cdot B_{i}(n) + O(\ar^2)
~~~~(\delta^m_i ~\mbox{is Kronecker symbol}),
\label{1.cf}
\eea
where the NLO coefficients $B_{i}(n)$ are exactly known (see, for example,
\cite{Buras}).

The $Q^2$-evolution of the moments $f_i(n,Q^2)$
is given by the well
known perturbative QCD \cite{Buras, Yndu} formulae:
\begin{eqnarray}
\frac{f_{NS}(n,Q^2)}{f_{NS}(n,Q_0^2)}&=&\bigg[\frac{\ar(Q_0^2)}
{\ar(Q^2)}\bigg]^{\gamma_{NS}^{(0)}(n)/2\beta_0}
\cdot H^{NS}(n,Q^2, Q_0^2)~~~, \nonumber \\
& & \nonumber \\
 f_j(n,Q^2) &=& f_j^+(n,Q^2) + f_j^-(n,Q^2) ~~~~~(j=S,G)
\label{3} \\
\frac{f_j^{\pm}(n,Q^2)}{f_j^{\pm}(n,Q_0^2)}&=&\bigg[\frac{\ar(Q_0^2)}
{\ar(Q^2)}\bigg]^{\gamma_{\pm}^{(0)}(n)/2\beta_0}
\cdot H_j^{\pm}(n,Q^2, Q_0^2)~~~, \nonumber 
\end{eqnarray}
where\footnote{We use a non-standard definition
(see \cite{YF93})
of the projectors $\ep_{jl}^{\pm}(n)$, which is very convenient beyond LO 
(see Eq. (\ref{3.4}) and \cite{Q2evo,HT}). The connection with the more 
usual definition $\alpha$, $\tilde \alpha$ and $\ep$ in ref.
\cite{Reya,Buras} is given by: $\ep_{SS}^{-}(n)= \alpha (n)$,
$\ep_{SG}^{-}(n)= \tilde \alpha (n)$ and $\ep_{GS}^{-}(n)= \ep (n)$ }
 \begin{eqnarray}
 f_j^{\pm}(n,Q^2) &=&
%\sum_{l=S,G}
\ep_{jl}^{\pm}(n) f_l(n,Q^2),~~~~~(j,l=S,G)
%d_{ab}=\frac{\gamma^{(0)}_{ab}(n)}{2\beta_0},
\label{3.1} \\
\gamma_{\pm}^{(0)}(n)&=& \frac{1}{2}
\Biggl[
\Bigl( \gamma_{GG}^{(0)}(n)+\gamma_{SS}^{(0)}(n) \Bigr)\pm
%\Bigl( \gamma_{SS}^{(0)}(n)- \gamma_{GG}^{(0)}(n) \Bigr)
\sqrt{ \Bigl( \gamma_{SS}^{(0)}(n)- \gamma_{GG}^{(0)}(n) \Bigr)
  +4\gamma_{GS}^{(0)}(n) \gamma_{SG}^{(0)} }
\Biggr] \nonumber \\
%\mbox{ and }~
 \ep_{qq}^{\pm}(n)&=&  \ep_{gg}^{\mp}(n) = \frac{1}{2}
\biggl( 1+ \frac{ \gamma_{SS}^{(0)}(n)- \gamma_{GG}^{(0)}(n)}
{\gamma_{\pm}^{(0)}(n)- \gamma_{\mp}^{(0)}(n)} \biggr),~~ 
  \ep_{jl}^{\pm}(n) = \frac{ \gamma_{jl}^{(0)}(n)}{\gamma_{\pm}^{(0)}(n)-
\gamma_{\mp}^{(0)}(n)} ~~~
(j \neq l)
\nonumber 
 \end{eqnarray}

The functions $H^{NS}(n, Q^2, Q^2_0)$ and $H_j^{\pm}(n, Q^2, Q^2_0)$  
%$$H_n(Q^2)=C_{NS}^{(n)}(a_s(Q^2)) \cdot \Gamma _{NS}^{(n)}(a_s(Q^2)) $$
are nonzero above the leading order (LO) approximation and may be 
represented as
 \begin{eqnarray}
H^{NS}(n, Q^2, Q^2_0) &=& 
1 + \Bigl( \ar(Q^2) - \ar(Q^2_0) \Bigr) Z_{NS}(n) + 
O \Bigl( \ar^2(Q^2) \Bigr) \nonumber \\ 
%\label{3.2} \\
& & \nonumber \\
H_j^{\pm}(n, Q^2, Q^2_0) &=& 
1 +  \Bigl( \ar(Q^2) - \ar(Q^2_0) \Bigr) Z_{\pm \pm}(n)
%\nonumber \\ 
\label{3.2} \\ &+& 
\Biggl( \ar(Q^2_0) \bigg[\frac{\ar(Q_0^2)}
{\ar(Q^2)}\bigg]^{\biggl(\gamma_{\mp}^{(0)}(n)-\gamma_{\pm}^{(0)}(n)\biggr)
/2\beta_0} 
%\nonumber \\&-& 
-\ar(Q^2) \Biggr)
Z^j_{\pm \mp}(n) +
O \Bigl( \ar^2(Q^2) \Bigr), 
\nonumber
 \end{eqnarray}
where
 \begin{eqnarray}
Z_{NS}(n) &=& \frac{1}{2\beta_0} \biggl( \gamma_{NS}^{(1)}(n) -
\gamma_{NS}^{(0)}(n) \cdot \frac{\beta_1}{\beta_0} \biggr) 
%\nonumber 
\label{3.21} \\
Z_{\pm \pm}(n) &=& \frac{1}{2\beta_0} \biggl( \gamma_{\pm \pm}^{(1)}(n) -
\gamma_{\pm}^{(0)}(n) \cdot \frac{\beta_1}{\beta_0} \biggr)
\label{3.3} \\
Z^S_{\pm \mp}(n) &=& \frac{\gamma_{\pm \mp}^{(1)}(n)}{2\beta_0 +
\gamma_{\pm}^{(0)}(n) - \gamma_{\mp}^{(0)}(n)}, ~~~
Z^G_{\pm \mp}(n) = Z^S_{\pm \mp}(n) \cdot \frac{\ep^{\mp}_{GG}}{\ep^{\mp}_{SS}}
%\nonumber 
\label{3.31}
 \end{eqnarray}
and
 \begin{eqnarray}
\gamma_{\pm\pm}^{(1)}(n)&=&  \sum_{j,l=S,G} \ep^{\pm}_{lj} \gamma^{(1)}_{jl}
%\nonumber \\
 \label{3.4} \\
\gamma_{\pm\mp}^{(1)}(n)&=&  \sum_{j=S,G} \ep^{\pm}_{jG} \gamma^{(1)}_{Gj} -
\ep^{\mp}_{SS} \gamma^{(1)}_{SS} +
\Bigl(\ep^{\pm}_{GS} - \ep^{\pm}_{GG}/\ep^{\pm}_{SG} \Bigr) \gamma^{(1)}_{GG} 
\nonumber 
\end{eqnarray}

As usually, here we use 
%$\beta_0$ and $\beta_1$,
$\gamma_{NS}^{(0)}(n)$, $\gamma_{jl}^{(0)}(n)$ $(j,l= S,G)$ and
$\gamma_{NS}^{(1)}(n)$, $\gamma_{jl}^{(1)}(n)$
as the first and the second terms 
%of the expansion 
with respect to $\ar $ of 
%QCD $\beta$-function and 
anomalous dimensions
$\gamma_{NS}(n,\ar)$ and $\gamma_{jl}(n,\ar)$ (see, for example, 
\cite{FlKoLa}).\\

{\bf 2.1.3}~~ 
In this subsection, we would like to discuss a possible
dependence of our results on the factorization scale $\mu_F$ and
the renormalization scale $\mu_R$, which appear (see, for example,
\cite{ViMi, scheme})
because perturbative series are
truncated. These scales 
%\bea
$\mu^2_F = k_F Q^2$
%,~~~ ~~~ 
and $\mu^2_R= k_R \mu^2_F = k_R k_F Q^2$
can be added to the r.h.s. of the equations
%modify the above equations 
(\ref{3.a}) and (\ref{3}), respectively. 

Then, the
equations (\ref{3.a}) are replaced by 
\begin{eqnarray}
M_n^{NS}(Q^2) &=& K_{NS}(f) \cdot
\hat C_{NS}^{tw2}(n, \ar(k_F Q^2)) \cdot f_{NS}(n,k_F Q^2) \nonumber \\
%\label{3.a} \\
& &  \label{3.al} \\
M_n^{S}(Q^2) &=& K_{S}(f) \cdot \biggl[
\hat C_{S}^{tw2}(n, \ar(k_F Q^2)) \cdot f_{S}(n,k_F Q^2)
+ \hat C_{G}^{tw2}(n, \ar(k_F Q^2)) \cdot f_{G}(n,k_F Q^2) \biggr]~.
\nonumber 
%\label{3.b}
\end{eqnarray}

The equations (\ref{3})  are replaced correspondingly by 
\begin{eqnarray}
\frac{f_{NS}(n, k_F Q^2)}{f_{NS}(n,k_F Q_0^2)}
&=&\bigg[\frac{\ar(k_F k_R Q_0^2)}
{\ar(k_F k_R Q^2)}\bigg]^{\gamma_{NS}^{(0)}(n)/2\beta_0}
\cdot \hat H^{NS}(n,k_F k_R Q^2, k_F k_R Q_0^2)~~~, \nonumber \\
& & 
\label{3l} \\
\frac{f_j^{\pm}(n,k_F Q^2)}{f_j^{\pm}(n,k_F Q_0^2)}
&=&\bigg[\frac{\ar(k_F k_R Q_0^2)}
{\ar(k_F k_R Q^2)}\bigg]^{\gamma_{\pm}^{(0)}(n)/2\beta_0}
\cdot \hat H_j^{\pm}(n,k_F k_R Q^2, k_F k_R Q_0^2)~. \nonumber 
\end{eqnarray}

The coefficients $\hat C_{NS}$, $\hat C_{S}$, $\hat C_{G}$, $\hat H^{NS}$  
and $\hat H_j^{\pm}$ can be obtained from the ones $C_{NS}$, $C_{S}$, 
$C_{G}$, $H^{NS}$ and $H_j^{\pm}$ by modification in the r.h.s. of
equations (\ref{1.cf}), (\ref{3.21}) and (\ref{3.3}) as follows:
%to the form:

in Eq. (\ref{1.cf})
\bea
\ar(Q^2) &\to& \ar(k_F Q^2) 
\label{2.cc0} \\
%\nonumber \\
B_{NS}(n) &\to& B_{NS}(n) + 
\frac{1}{2}\gamma_{NS}^{(0)}(n) \ln{k_F} 
\nonumber \\
B_{j}(n) &\to& B_{j}(n) + \frac{1}{2}\gamma_{jS}^{(0)}(n) \ln{k_F} ~~~~~~
 (j=S,G)  
\label{2.cc1} 
\eea

in Eqs. (\ref{3.21}) and (\ref{3.3})
\bea
\ar(Q^2) &\to& \ar(k_F k_R Q^2) 
\label{2.cf0} \\
Z_{NS}(n) &\to& Z_{NS}(n) + \frac{1}{2}\gamma_{NS}^{(0)}(n)
\ln{k_R} 
%~~~~~~~~~~ ~~~~~~~~~~~~ \mbox{ in Eq.(\ref{3.21})}
\nonumber \\
Z_{\pm\pm}(n) &\to& Z_{\pm\pm}(n) + \frac{1}{2} \gamma^{(0)}_{\pm}(n)
\ln{k_R} 
%~~~~~~~~~~ ~~~~~~~~~~~\,\, \mbox{ in Eq.(\ref{3.3})}
\label{2.cf}
\eea

The Eqs. (\ref{2.cc1}) can be obtained easily using, for example,  
the results of \cite{KaKo}.
The Eqs. (\ref{2.cf}) 
can be found from the expansion of the coupling constant $\ar(k_F k_R Q^2)$
around the one $\ar(Q^2)$ in the r.h.s. of the exact
%obey to the 
solution of DGLAP equations (see Eqs.(\ref{3l}) and (\ref{3.2})).

The changes (\ref{2.cc1}) and (\ref{2.cf}) of the results for 
$Q^2$-dependence under variation 
of $k_F$ and $k_R$ (usually 
\footnote{In the recent articles \cite{BlNe,NeVo,NeVo1,KPS} the variation
from $1/4$ to $4$ has been used. In our opinion, the case
$k_F=k_R=4$ leads to very small scale of coupling constant: $Q^2/16$,
that requires to reject many of experimental points of used data,
because we have the general cut $Q^2 > 1$ GeV$^2$. So, we prefer to use
the variation of scales from $1/2$ to $2$.}
from $1/2$ to $2$) give an estimation of the
errors due to renormalization and factorization scale uncertainties. 
Evidently that, by definition, these uncertainties are connected with
the impact of unaccounted terms of the perturbative series
and can represent {\it theoretical uncertainties} in values of fitted 
variables.
Indeed, an incorporation of NNLO corrections to the analysis strongly 
suppress these uncertainties (see \cite{NeVo,NeVo1}).

We study exactly the $\mu_F$ and $\mu_R$ dependences here for fitted 
values of coupling constant. 
The results
of the study are given in the Section 6.

As one can see in Eqs. (\ref{2.cc0}) and (\ref{2.cf0}), the coupling
constant $\ar$ has different arguments in the NLO corrections  
%and $\hat C_{G}$, 
of coefficient functions $\hat C_{NS}$ and $\hat C_{j}$ ($j=S,G$)
and in the NLO corrections $\hat H^{NS}$  
and $\hat H_j^{\pm}$ of the $Q^2$-evolution of parton
distributions. We would like to note that the difference between the
corresponding coupling constants $\ar(k_FQ^2)$ and $\ar(k_F k_R Q^2)$
is proportional to $\ar^2$ and, thus, mathematically negligible in our
NLO approximation.\\
Then, we can use the replacement
%change 
(\ref{2.cf0}) in coefficient functions too,
as it has been done in previous studies \cite{BlNe,NeVo,NeVo1,KPS1}.
We note that 
the replacement $\ar(k_FQ^2) \to \ar(k_F k_R Q^2)$ in Eq. (\ref{2.cc0})
increases slightly
%essentially 
the factorization scheme dependence of the results for
coupling constant (see analyses based on nonsinglet evolution 
and discussions in Section 6).

\subsection{ Normalization of parton distributions }

The moments $f_i(n,Q^2)$ at some $Q^2_0$ is theoretical input of our analysis
which is  fixed as follows.\\

For 
%nonsinglet 
fits of data at $x\geq 0.25$ we can work only with the nonsinglet parton 
density and use
directly its normalization $\tilde f_{NS}(x,Q_0^2)$ (see, for example,
\cite{PKK}-\cite{KPS}):
\bea
f_{NS}(n,Q_0^2) &=& \int_0^1 dx x^{n-2} \tilde f_{NS}(x,Q_0^2), \nonumber \\
\tilde f_{NS}(x,Q_0^2) &=& A_{NS}(Q_0^2)
%x^{b_{NS}(Q_0^2)}
(1-x)^{b_{NS}(Q_0^2)}
(1+d_{NS}(Q_0^2)x)
\label{4}
\eea
where $A_{NS}(Q_0^2)$, 
%$b_{NS}(Q_0^2)$, 
$b_{NS}(Q_0^2)$ and $d_{NS}(Q_0^2)$ are some coefficients
\footnote{We do not consider here the term $\sim x^{a_{NS}(Q_0^2)}$
in the normalization $\tilde f_{NS}(x,Q_0^2)$, because $x\geq 0.25$.
The correct small-$x$ asymptotics of nonsinglet distributions will
be obtained by Eq.(\ref{4.2d1}) from the corresponding parameters of
the valent quark distributions (\ref{4.2v}) fitted with complete
singlet and nonsinglet evolution in Section 5.}.\\

At the analyses at arbitrary values of $x$
%singlet case 
we should introduce the normalizations for densities of
individual quarks $(q=u,d,s,...)$ and antiquarks 
$(\overline q=\overline u,\overline d,\overline s,...)$
$\tilde f_{q}(x,Q_0^2)$ and $\tilde f_{\overline q}(x,Q_0^2)$ having the 
moments:
\bea
f_{i}(n,Q_0^2) ~=~ \int_0^1 dx x^{n-2} \tilde f_{i}(x,Q_0^2),
\label{4.1}
\eea

The distributions of $u$ and $d$ quarks 
$\tilde f_{u}(x,Q_0^2) \equiv u(x,Q_0^2)$ and  
$\tilde f_{d}(x,Q_0^2) \equiv d(x,Q_0^2)$ are split in two components:
the valent one $u_v(x,Q_0^2)$ and $d_v(x,Q_0^2)$ and the sea one
$u_v(x,Q_0^2)$ and $d_v(x,Q_0^2)$. For other quark distributions and
antiquark densities we keep only sea parts. Moreover, following 
\cite{Buras,CDR}
we suppose equality of all sea parts and mark their sum as $S(x,Q_0^2)$.

We use the following parameterizations for densities 
$u_{v}(x,Q_0^2)$, $u_{v}(x,Q_0^2)$, $S(x,Q_0^2)$
% $\tilde f_{\overline q}(x,Q_0^2)$ and
$\tilde f_{G}(x,Q_0^2)$:
\bea
u_v(x,Q_0^2) &=& \frac{2}{B(a_u(Q_0^2),b_u(Q_0^2)+1)}
\,x^{a_u(Q_0^2)}(1-x)^{b_u(Q_0^2)},  \nonumber \\
d_v(x,Q_0^2) &=& \frac{1}{B(a_d(Q_0^2),b_d(Q_0^2)+1)}
\,x^{a_d(Q_0^2)}(1-x)^{b_d(Q_0^2)},  
\label{4.2v}\\
%\nonumber \\
S(x,Q_0^2) &=& C_S(Q_0^2)x^{a_S(Q^2_0)} (1-x)^{b_S(Q_0^2)},~~~~~~ 
\nonumber \\
\tilde f_{G}(x,Q_0^2) &=& C_{G}(Q_0^2)x^{a_G(Q^2_0)} (1-x)^{b_{G}(Q_0^2)},
\label{4.2}
\eea
where
$B(a,b)$ is the Euler beta-function.
The parameterizations (\ref{4.2v}) have been chosen to satisfy 
(at the normalization point $Q^2_0$) the known rule:
\bea
 \int_0^1 dx V(x,Q^2)=3, 
%~~~~~~~~ (V(x,Q^2)=u_v(x,Q^2)+d_v(x,Q^2))
  \nonumber 
\eea
where $V(x,Q^2)=u_v(x,Q^2)+d_v(x,Q^2)$ is the distribution of valent quarks.

We 
%would like to 
note that the
nonsinglet and singlet parts of quark distributions
$\tilde f_{NS}(x,Q_0^2)$ and $\tilde f_{S}(x,Q_0^2)$ 
can be represented as combination of quark ones
\bea
\tilde f_{S}(x,Q_0^2) &\equiv& \sum_{q}^{f}\tilde f_{q}(x,Q_0^2)
~=~ 
%u_v(x,Q_0^2) + d_v(x,Q_0^2) 
V(x,Q_0^2) + S(x,Q_0^2)
\label{4.2d}\\
\tilde f_{NS}(x,Q_0^2) &=& u_v(x,Q_0^2) - d_v(x,Q_0^2),
\label{4.2d1}
\eea
where the r.h.s. of
Eq.(\ref{4.2d1}) is correct only in the framework of our supposition about
equality of antiquarks distributions and sea components of quark ones.

In principle, following the PDF models used in \cite{fits,Al2000} 
and above Eq.(\ref{4}) one
can add in Eq.(\ref{4.2}) terms proportional to $\sqrt{x}$ and $x$. 
However, 
%since these 
the terms $\sim \sqrt{x}$
are important only in the region of rather small $x$
(see discussion in \cite{Al2000}). The terms $\sim x$ lead only to 
%changings
replacement
of $C_i$, $a_i$ and $b_i$ values (see, for example, \cite{VoKoMa}). Thus, 
we neglect these terms in our
analysis.

In the most our fits
we do not take into account also the terms $\sim x^{a_G(Q^2_0)}$ and
$\sim x^{a_S(Q^2_0)}$ into gluon and sea quark distributions, because
we do not consider experimental data at small values of Bjorken variable 
$x$ \footnote{However, we have performed several fits with nonzero 
$a_G$ and $a_S$ values taken into account (see Section 5). We have 
found a negative value
of them: $a_G = a_S \sim -0.18$ (that is in agreement with \cite{Al2000})
but these results cannot be considered
seriously without taking into account H1 and ZEUS data \cite{H1,ZEUS}
(see, however, discussions in the subsection 5.3.4).}.
We hope to include H1 and ZEUS data \cite{H1,ZEUS} in our future 
investigations \cite{KKrenorm} and then
to study $Q^2$-dependence of the coefficients $a_G(Q^2)$ and $a_S(Q^2)$,
which could be very nontrivial (see, for example, Refs.
\cite{AbraLevin,JeKoPa,H1,Q2evo} and references therein).\\

We impose also the condition for full momentum conservation in the form:
\bea
1~=~P_{G}(Q^2) + P_{q}(Q^2),
\label{4.3}
\eea
where  
\bea
P_{G}(Q^2) &=& \int_0^1 dx \tilde f_{G}(x,Q^2),  \nonumber \\
P_{q}(Q^2) &=& \int_0^1 dx \biggl(\tilde f_{NS}(x,Q^2) + 
\tilde f_{S}(x,Q^2)\biggr),
\label{4.3a}
\eea

The coefficients $C_i(Q_0^2)$, $a_i(Q_0^2)$, $b_i(Q_0^2)$, $c_i(Q_0^2)$ and
$d_i(Q_0^2)$ should be found together with $h_4(x)$
%, $h_6(x)$ 
(see subsection 2.6) and 
the normalization $\alpha_s(M^2_Z)$ of QCD coupling constant (or QCD
parameter $\Lambda $) by the fits of experimental data.\\

\subsection{ Target mass corrections }

%{\bf 2.}~
The target mass corrections \cite{GP,Buras} modify the SF
$F_2^{NS}(x,Q^2)$ in the following way
\begin{eqnarray}
F_{2}^{pQCD}(x,Q^2) 
&=&
\frac{1}{r^3}\frac{x^2}{ \xi^2} F_2^{tw2}(\xi,Q^2)
+6\frac{M_{nucl.}^2}{Q^2} \frac{x^3}{r^6} \int^1_{\xi}
\frac{d\xi'}{(\xi')^2}F_2^{tw2}(\xi',Q^2)
\nonumber \\
&+& 12\frac{M_{nucl.}^4}{Q^4} \frac{x^4}{r^5} \int^1_{\xi} \int^1_{\xi'}
\frac{d\xi''}{(\xi'')^2}F_2^{tw2}(\xi'',Q^2)
\label{4.1}
\end{eqnarray}
where 
%$A_2^{'}$ is the free parameter, 
$M_{nucl}$ is the mass of the nucleon, $r=\sqrt{1+ x^2 M_{nucl.}^2/Q^2}$
and the Nachtmann variable $\xi = 2x/(1+r)$.

In our analyses below,
%Here 
we will use all this (\ref{4.1})
%, (\ref{4.2}), (\ref{4.3})
representation 
\footnote{It is contrary to \cite{Al2001}, where
only the term $\sim M_{nucl.}^2/Q^2$ has been used. We note that 
the appearance of the terms $\sim M_{nucl.}^2/Q^2$ at $x=1$ (see, for example, 
\cite{TMC}), i.e. the absence of the equality $F_2^{pQCD}(1,Q^2)=0$,
is not important in the our analyses because we do not use experimental
data at very large $x$ values: $x \leq 1$.}.
We would like to keep the full value of kinematic power corrections,
given by nonzero nucleon mass. Then, the excess of $1/Q^2$ dependence
encoded in experimental data will give the magnitude of twist-four
corrections, which is most important part of
dynamical power corrections.

\subsection{ Thresholds of heavy quarks }

Modern estimates performed in ~\cite{yuri, gotts}
have revealed a quite significant role of threshold effects in the
 $\as(Q^2)$ evolution when the DIS data lie close to
threshold points $Q^2=M_{f+1}^2 \sim m_{f+1}^2$ (to the position of 
so-called "Euclidean--reflected" threshold of heavy particles).
The corresponding corrections to the normalization
$\asMZ$ can reach several percent, i.e. , they are of the order of 
other uncertainties which should be under control at our analysis.

An appropriate procedure for the inclusion of threshold effects into the
$Q^2$--dependence of $\as(Q^2)$ in the framework of the massless
$\MSbar$ scheme was proposed more than 10 years ago ~\cite{bern,marc} :
transition from the region with a given number of flavors $f$
described by massless $\as(Q^2;f)$
\footnote{Following \cite{SSiMi1} 
in this subsection we use the form $\as(Q^2;f)$
for the coupling constant with purpose to demonstrate  its $f$-dependence
through the ones of $\beta_0$ and $\beta_1$ coefficients.} 
to the
next one with $f + 1$ (``transition across the $M_{f+1}$ threshold")
is realized here with the use of the so--called ``matching relation"
for $\as(Q^2)$ ~\cite{marc}.  The latter may be considered as the
continuity condition for $\as(Q^2)$ on (every) heavy quark mass $m_{f+1}$
\bea 
 \as(Q^2=M^2_{f+1}; f)&=& \as(Q^2=M^2_{f+1}; f+1)~~~~~~~~
\mbox{ and } \label{h1}\\ 
& & \nonumber \\
M_{f+1}&=& m_{f+1}
\label{t1}
\eea
that provides an accurate $\as(Q^2)$--evolution description for $Q^2$
values not close to the threshold region (see \cite{SSiMi1} and references 
therein). 

At the analyses based on nonsinglet evolution, 
the additional $f$-dependence comes only
from the NLO correction of NS anomalous dimensions (see \cite{FlKoLa})
\footnote{The corresponding 
 moments at any $Q^2$ value are proportional to same 
coefficient $K_{NS}(f)$. Thus, the coefficient can be always taken up by
the normalization $M_n^{NS}(Q^2_0)$.}. In the Section 3  we check
numerically the dependence of the results from the matching point. We use
two matching points: (\ref{t1}) one and 
%(\ref{t1}) one  
\bea
M_{f+1}~=~2 \,m_{f+1}
\label{t2.0}
\eea
and
demonstrate very little variations of the results
\footnote{We will not take into account a small variation (see \cite{RodSan})
of the
continuity condition (\ref{h1}) because of the matching point (\ref{t2.0}).}
(see Section 4 and discussion there).

As we know, for singlet part of evolution no 
simple recipe exists for exact value of the matching point $M_{f+1}$. From
one side, as in the nonsinglet case, there is $Q^2$-evolution of the SF
moments which leads to above condition (\ref{t1}). But here we have also
the generation of heavy quarks (at lowest nontrivial order, in the framework 
of the
photon-gluon fusion process), that gives contributions to gluon part
of the singlet coefficient function. The  photon-gluon fusion needs the
matching point at the value of $Q^2$, when $W^2=4m_{f+1}^2$, i.e.
\bea
M_{f+1}^2 \cdot \frac{1-x}{x}+M^2_{nucl.} ~=~4m_{f+1}^2
\label{t2}
\eea

At small $x$ values the condition (\ref{t2}) is quite close to the one
(\ref{t1}) (for example, at $x=0.2$ $M_{f+1}^2~=~m_{f+1}^2 - M^2_{nucl.}/4$),
but at the range of large and intermediate values of $x$, the value of
$M_{f+1}$ is essentially large to compare with one of Eq.(\ref{t1}).
For example, at $x=0.5$ $M_{f+1}^2~=~4m_{f+1}^2 - M^2_{nucl.}$, that
is very close to the matching point (\ref{t2.0}). At larger $x$ values
the value of $M_{f+1}^2$ will be close to ones in \cite{BlNe,NeVo}.

We note, that the difference between nonsinglet and singlet $Q^2$-dependences
comes from contribution of gluon distribution. The contribution is
negligible at $x> 0.3$ that supports qualitatively the choice (\ref{t1}) as  
the matching point.

We would like to note also, that at NLO approximation and above,
%In general, 
the situation is even more difficult in singlet case, because every 
subprocess 
%(at $\alpha_s^2$ level and beyond)
generates itself matching point $M_{f+1}$ to coefficient functions.
To estimate a possible effect of a dependence on matching point,
we will fit data (in Section 3) with two different matching points: 
(\ref{t1}) one and (\ref{t2.0}) one. Surprisingly,
at the singlet case, where all functions coming to $Q^2$-evolution are 
$f$-dependent, we do not find a strong $f$-dependence of our results
(see Section 5 and discussions there).

\subsection{ Nuclear effects }

Starting with EMC discovery in \cite{EMCr}, it is well known about
the difference between PDF in free hadrons and ones in hadrons in nuclei.
We incorporate the difference in our analyses.

In the nonsinglet case we parameterize the initial PDF in the form 
%closed to 
(\ref{4}) for every type of target. We have

%\begin{equation}
\bea
f^A_{NS}(n,Q_0^2)~=~\int_0^1 dx x^{n-2} \tilde f^A_{NS}(x,Q_0^2),
\label{nu1}
\eea
where
\bea
\tilde f^A_{NS}(x,Q_0^2)~=~
A^A_{NS}(Q_0^2)
%x^{b^A_{NS}(Q_0^2)}
(1-x)^{b^A_{NS}(Q_0^2)}
(1+d^A_{NS}(Q_0^2)x)
\label{nu2}
%\end{equation}
\eea
and
$A=H,~D,~C$ and $F$ in the case of $H_2,~D_2,~C^{12}$ and $Fe^{56}$ targets,
respectively.

In the singlet case we have many parameters in our fits, which should be 
fitted very carefully. 
%Thus, 
The representations similar to (\ref{nu2}) for gluon and sea quark PDF  
should complicate 
%so much 
our analyses. To overcome the 
%complicity 
problem,
we apply the Eqs. (\ref{nu1}) and (\ref{nu2}) only to $H_2$ and $D_2$ cases. 
For heavier
targets we apply simpler representations for $F_2^A$ structure functions
in the form:
\bea
F^A_{2}(x,Q^2) ~=~ F^D_{2}(x,Q^2) \cdot
K^A_{1}
\Bigl(1-K^A_{2}x +K^A_{3}x^2 \Bigr)~~~~~(A=C^{12}, Fe^{56}),
\label{nu3}
%\end{equation}
\eea
where we use experimental observation 
\footnote{The small $Q^2$ dependence of EMC ration has been observed also
in theoretical studies. For example, in the framework of rescaling model
\cite{rescaling} the $Q^2$ dependence is very small (see \cite{AVKEMC}). 
It has double-logarithmic form
and locates only in argument of Euler $\Psi$-function.}
(see \cite{BCDMSr} and references 
therein) about approximate $Q^2$-independence of EMC ration 
$F^A_{2}/F^D_{2}$.

%\vskip 1cm

\subsection{ Higher-twist corrections }

%{\bf 3.}~
For $n$-space 
 eq.(\ref{1}) transforms to
\begin{equation}
M_n^{full}(Q^2)~=~ M_n^{pQCD}(Q^2) + 
%\biggl[
\frac{h_4(n)}{Q^2}, 
\label{1.1}
\end{equation}
where 
%(hereafter ($a=4, 6$))
$ h_4(n, Q^2)$ are the moments of the function $ \tilde h_4(x, Q^2)$:
\begin{equation}
h_4(n) = \int_0^1 x^{n-2} \tilde h_4(x)dx
%/M_n^{NS}(Q^2)
\label{1.11}
\end{equation}

%{\bf 2.}~
The shape $\tilde h_4(x)$ (or coefficients $ h_4(n)$)  of the twist-four
%(HT) 
corrections  are of primary consideration in our
analysis. They can be chosen in the several different forms:

\begin{itemize}

\item 
%Here 
the twist-four terms (and twist-six ones) are fixed in agreement with the 
infrared 
renormalon (IRR) model (see \cite{DW,SMMS,Maul,Beneke} and references
therein).
\item 
The twist-four 
term in the form 
$ h_4(x)\sim \frac{d}{dx}lnF_2^{NS}(x,Q^2) \sim 1/(1-x) $ 
(see \cite{Yndu} and references therein).
This behavior matches 
%(see \cite{CDR}) 
the fact that higher twist effects are usually important only at higher $x$. 
The twist-four coefficient function has the form
%\begin{eqnarray}
$
%C_4(n) =
C_4^{der}(n)
= (n -1) A_4^{der}$.
\item
The twist-four term $\tilde h_4(x)$
is considered as a set free parameters at each $x_i$ bin. The set has the form
$\tilde h_4^{free}(x)=\sum_{i=1}^{I} \tilde h_4(x_i)$, 
where $I$ is the number of bins.
The
constants $\tilde h_4(x_i)$ (one per $x$-bin) parameterize $x$-dependence of 
$\tilde h_4^{free}(x)$.
\end{itemize}

The first two cases have been considered already in \cite{KKDIS} and
will be studied carefully later \cite{KKrenorm}. Here we will follow
the last possibility \footnote{In conclusion we present, however, several 
comments 
%and consider a 
about an application of higher-twist corrections in the form of IRR
%renormalon 
model.}.

%%%%%%%%%%%%%%%%  3  3  3  3 %%%%%%%%%%%%%%%%%%%

\section{ Fits of $F_2$: procedure }

To clear up the importance of HT terms we fit 
%SLAC and BCDMS  ($H_2$ and $D_2$) data \cite{SLAC1}-\cite{BCDMS2} 
SLAC, NMC,  BCDMS and BFP
experimental data \cite{SLAC1}-\cite{BFP}
(including the it systematic
errors),
keeping identical form of perturbative part
at NLO approximation.
In the Section we demonstrate the basic ingredients of the analyses.

As it has been already discussed in the Introduction
we use the exact solution of DGLAP equation
for the Mellin moments $M_n^{tw2}(Q^2)$ (\ref{2}) of
%$F_2^{full}(x,Q^2)$, $F_2^{pQCD}(x,Q^2)$ and
SF $F_2^{tw2}(x,Q^2)$
and
the subsequent reproduction of $F_2^{full}(x,Q^2)$, $F_2^{pQCD}(x,Q^2)$  and/or
$F_2^{tw2}(x,Q^2)$ at every needed $Q^2$-value with help of the Jacobi 
Polynomial expansion method. 
The method of the Jacobi polynomial expansion 
was developed
in \cite{Barker, Kri} and described in details
in Refs.\cite{Kri}. Here we consider only
some basical definitions.

Having the QCD expressions for the Mellin moments
$M_n^{k}(Q^2)$ we can reconstruct the SF $F_2^k(x,Q^2)$
%using the Jacobi polynomial expansion method:
as
\begin{equation}
F_{2}^{k,N_{max}}(x,Q^2)=x^{a}(1-x)^{b}\sum_{n=0}^{N_{max}}
\Theta_n ^{a , b}
(x)\sum_{j=0}^{n}c_{j}^{(n)}{(\alpha ,\beta )}
M_{j+2}^{k} \left ( Q^{2}\right ),
\label{2.1}
\end{equation}
where $\Theta_n^{a,}$ are the Jacobi polynomials
\footnote{We would like to note here that there is similar method 
\cite{Ynd}, based on Bernstein polynomials. The method has been used 
in the analyses at the NLO level in \cite{KaKoYaF,KaKo}
and at the NNLO level in \cite{SaYnd,KPS1}.}
and $a,b$ are the parameters, fitted by the condition
of the requirement of the minimization of the error of the
reconstruction of the
structure functions 
\footnote{ There is another possibility to fit 
%experimental 
data. It is
possible to transfer experimental information about structure functions
to their moments and to analyze directly these moments. This approach was 
very popular in the past (see, for example,
 Ref. \cite{DuRo}) but it is used very rarely
at present (see, however, \cite{Lyk} and references therein) because
a transformation of experimental information to the SF moments is quite
a difficult procedure.} (see Ref.\cite{Kri} for details).\\

First of all, we choose the cut $Q^2 \geq 1$ GeV$^2$ in all our studies.
For $Q^2 < 1$ GeV$^2$, the applicability of twist expansion is very
questionable. 

Secondly, we
choose quite large values of the normalization point
$Q^2_0$. There are several reasons of this choice:
\begin{itemize}
\item 
Our above perturbative formulae should be applicable at the value of
$Q^2_0$. Moreover, the higher order corrections $\sim \as^n(Q^2_0)$ 
($n \geq 2$), coming from normalization conditions of PDF,
 are less important at higher $Q^2_0$ values.
\item
It is necessary to cross heavy quark thresholds less number of time
to reach $Q^2=M^2_Z$, the point of QCD coupling constant normalization.
\item
It is better to have the value of $Q^2_0$ around the middle point
of {\it logarithmical} range of considered $Q^2$ values. 
%At least,
Then
at the case the higher order corrections $\sim (\as(Q^2)-\as(Q^2_0))^n$ 
($n \geq 2$) are less important.
\end{itemize}

%\vspace{0.5cm}

Basic characteristics of the quality of the fits are
$\chi^2/DOF$ for SF $F_2$ and for its slope $dlnF_2/dlnQ^2$, which has
very sensitive 
perturbative properties (see \cite{Buras}). 

As these fits involve many free parameters independent of perturbative
QCD, it is important to check whether, in the results of the fits, the 
features most specific to perturbative QCD are in good agreement with
the data. 
The slope $dlnF_2/dlnQ^2$ has really very sensitive 
perturbative properties and will be used (see the Figs. 4-8 and 10-12)
to check properties of fits.
Indeed, the DGLAP equations predict that the logarithmical derivations
of SF and PDF logarithms are proportional 
%exactly 
very nearly to coupling constant $\as(Q^2)$
with an $x$-dependent proportionality coefficient that depends 
(at $x > 0.2$) only
weakly on the $x$-dependence of the SF and PDF.
Thus, the study of the $Q^2$-dependence of the slope $dlnF_2/dlnQ^2$
leads to obtain the direct information about the corresponding 
$Q^2$-dependence of QCD coupling constant and to verify the range of 
accuracy for formulae of perturbative QCD.\\

We use MINUIT program \cite{MINUIT} for
minimization of two $\chi^2 $ values:
\begin{eqnarray}
\chi^2(F_2) = {\biggl|\frac{F_2^{exp} - F_2^{teor}}{\Delta F_2^{exp}}
\biggr| }^2 ~ \mbox{ and }  ~
\chi^2(\mbox{slope}) = {\biggl|\frac{D^{exp} - D^{teor}}{\Delta D^{exp}}
\biggr| }^2~~~~ \biggl(
D=\frac{dlnF_2}{dlnQ^2} \biggr)
\nonumber
\end{eqnarray}
%where $b=dlnF_2/dlnQ^2$.

We would like to apply the following procedure: we study the dependence of 
$\chi^2/DOF$ value on value of $Q^2$ cuts for various sets of experimental
data. The study will be done for the both cases: including higher twists
%terms 
corrections (HTC) and without them.

We use free normalizations of data for different experiments. 
For the reference, we use the most stable deuterium BCDMS data
at the value of energy $E_0=200$ GeV 
\footnote{$E_0$ is the initial energy lepton beam.}. 
%The usage 
Using other types of data as reference gives
negligible changes in our results. The usage of fixed normalization
for all data leads to fits with a bit worser $\chi^2$.

\section{ Results of fits of $F_2$: the nonsinglet evolution part  }

Firstly, we will consider the $Q^2$-evolution of the SF $F_2$ at the
nonsinglet case where there are the contributions of quark
densities only and, thus, the corresponding fits are essentially
simpler. 
The consideration of the nonsinglet part limits the range of data
by the cut $x \geq 0.25$. At smaller $x$-values the contributions
of gluon distribution is not already negligible.

Hereafter at nonsinglet case of evolution we choose
 $Q^2_0$ = 90 GeV$^2$ for the BCDMS data and the combined all data 
 and $Q^2_0$ = 20 GeV$^2$ for the combined SLAC, NMC, BFP one,
 respectively.
 The choice of $Q^2_0$-values  is in good agreement with above 
conditions (see the previous Section). We use also $N_{max} =8$,
the cut $0.25 \leq x \leq 0.8$.
%and $I=6$.
The $N_{max}$-dependence of the results has been studied carefully in
Ref. \cite{Kri} (see also below the Table 3).

\subsection { BCDMS $C^{12} + H_2 + D_2$ data }
%{\bf 4.1.} {\bf SLAC data}

We start our analysis with the most precise experimental data 
\cite{BCDMS1,BCDMS2,BCDMS3} obtained  by BCDMS muon
%$\mu h$ 
scattering experiment at the high $Q^2$ values.
The full set of data is 607 points (when $x \geq 0.25$).
The starting point of QCD evolution is $Q^2_0=90$ GeV$^2$.

It is well known that the original analyses 
%of 
given by BCDMS Collaboration itself (see
also Ref. \cite{ViMi}) lead to quite small values of $\alpha_s(M^2_Z)$:
for example, $\alpha_s(M^2_Z)=0.113$ has been obtained in \cite{ViMi}
\footnote{ We would like to note that the paper \cite{ViMi} has a quite
strange result. Authors of the article have obtained the value 
$\Lambda^{(4)}_{\MSbar}=263$ MeV, that should lead to the value of
coupling constant $\alpha_s(M^2_Z)$ is equal to $0.1157$.}.
Although in some recent papers (see, for example, 
\cite{Al2000,Al2001,H1BCDMS})
more higher values of $\alpha_s(M^2_Z)$ have been observed, we think that
an additional reanalysis of BCDMS data should be very useful. 

Based on study \cite{Kri2} (see also \cite{Syst,H1BCDMS}) we propose that 
the reason for small values
of $\alpha_s(M^2_Z)$ coming from BCDMS data is the existence of the subset
of the data having large systematic errors. Indeed, the original analyses
of $H_2$, $D_2$ and $C^{12}$ data performed by BCDMS Collaboration lead
to the following value of QCD mass parameter (see Refs. 
\cite{BCDMS1,BCDMS2,BCDMS3}:
\bea
\Lambda^{(4)}_{\MSbar} ~=~ 
\biggl( 220 \pm 13  ~\mbox{(stat)} 
\pm 50 ~\mbox{(syst)} \biggr)\, \mbox{MeV},
\label{BCDMS}
\eea
i.e. the systematic error is four times bigger than the statistical one.
Hereafter the symbols ``stat'' and ``syst'' 
%and ``norm'' 
mark the statistical error and
 systematic one, 
%and the error of normalization of experimental data,
respectively.

We study this subject by 
introducing several so-called $Y$-cuts 
\footnote{Hereafter we use the kinematical variable $Y=(E_0-E)/E_0$,
where $E_0$ and $E$ are initial and scattering energies of lepton, 
respectively.}
(see \cite{Kri2} and subsections
4.1.1 and 5.1.1). Excluding this set of data with large systematic errors
leads to essentially larger values of $\alpha_s(M^2_Z)$ and very slow
dependence of the values on the concrete choice of the $Y$-cut (see below).\\

{\bf 4.1.1. The study of systematics. }\\
%{\bf Nonsinglet case}

The correlated systematic errors of the data have been studied in \cite{Kri2},
together with the other parameters. Regions of data have been identified
in which the fits cause large systematic shifts of the data points. We 
would like to exclude these regions from our analyses.

We have studied influence of the experimental systematic errors on the
results of the QCD analysis as a function of $Y_{cut3}$, $Y_{cut4}$ and
$Y_{cut5}$ applied to the data.
We use the following $x$-dependent $y$-cuts:
\bea
& &y \geq 0.14 \,~~~\mbox{ when }~~~ 0.3 < x \leq 0.4 \nonumber \\
& &y \geq 0.16 \,~~~\mbox{ when }~~~ 0.4 < x \leq 0.5 \nonumber \\
& &y \geq Y_{cut3} ~~~\mbox{ when }~~~ 0.5 < x \leq 0.6 \nonumber \\
& &y \geq Y_{cut4} ~~~\mbox{ when }~~~ 0.6 < x \leq 0.7 \nonumber \\
& &y \geq Y_{cut5} ~~~\mbox{ when }~~~ 0.7 < x \leq 0.8 
\label{cut}
\eea

We use several sets $N$ of the values for the cuts at $0.5 < x \leq 0.8$, 
which are given in the Table 1.\\

{\bf Table 1.} The values of $Y_{cut3}$, $Y_{cut4}$ and $Y_{cut5}$.
\vspace{0.2cm}
\begin{center}
%\footnotesize
\begin{tabular}{|c|c|c|c|c|c|c|c|}
\hline
& & & & & & & \\
$N$ & 0 & 1 & 2 & 3 & 4 & 5 & 6 \\
& & & & & & & \\
\hline \hline
$Y_{cut3}$ & 0 & 0.14 & 0.16 & 0.16 & 0.18 & 0.22 & 0.23 \\  
%\hline
$Y_{cut4}$ & 0 & 0.16 & 0.18 & 0.20 & 0.20 & 0.23 & 0.24 \\
$Y_{cut5}$ & 0 & 0.20 & 0.20 & 0.22 & 0.22 & 0.24 & 0.25 \\
\hline
\end{tabular}
\end{center}

\vspace{0.5cm}

The systematic errors for BCDMS data are given \cite{BCDMS1,BCDMS2,BCDMS3}
as multiplicative factors to be applied to $F_2(x,Q^2)$: $f_r, f_b, f_s, f_d$
and $f_h$ are the uncertainties due to spectrometer resolution, beam momentum,
calibration, spectrometer magnetic field calibration, detector inefficiencies 
and energy normalization, respectively.

For this study each experimental point of the undistorted set was multiplied
by a factor characterizing a given type
of uncertainties and a new (distorted) data set was fitted again
in agreement with our procedure considered in the previous section. The factors
($f_r, f_b, f_s, f_d, f_h$) were taken from papers \cite{BCDMS1,BCDMS2,BCDMS3}
(see CERN preprint versions in \cite{BCDMS1,BCDMS2,BCDMS3}).
The absolute differences between the values of $\alpha_s$ for the distorted
and undistorted sets of data are given in Table 2 and 
the Fig. 1 as the total systematic
error of $\alpha_s$ estimated in quadratures. The number of the experimental
points and the value of $\alpha_s$ for the undistorted set of $F_2$ are also
given in the Table 2 and the Fig. 1. \\

{\bf Table 2.} The values of $\asMZ$ at different values of $N$.
\vspace{0.2cm}
\begin{center}
%\footnotesize
\begin{tabular}{|l|c|c|c|c|}
%\begin{tabular}{|p{35pt}|p{38pt}|p{38pt}|p{38pt}|p{38pt}|p{38pt}|p{38pt}|}
\hline
& & & &  \\
$N$  & number & $\chi^2(F_2)/DOF$  & $\as(90~\mbox{GeV}^2)$  $\pm$ stat 
&full  \\
&of points &  & &syst. error \\
\hline \hline
0 & 607 & 1.03 &  0.1590 $\pm$ 0.0020 & 0.0090 \\
1 & 511 & 0.97 &  0.1711 $\pm$ 0.0027 & 0.0075 \\
2 & 502 & 0.97 &  0.1720 $\pm$ 0.0027 & 0.0071 \\
3 & 495 & 0.97 &  0.1723 $\pm$ 0.0027 & 0.0063 \\
4 & 489 & 0.94 &  0.1741 $\pm$ 0.0027 & 0.0061 \\
5 & 458 & 0.95 &  0.1730 $\pm$ 0.0028 & 0.0052 \\
6 & 452 & 0.95 &  0.1737 $\pm$ 0.0029 & 0.0050 \\
\hline
\end{tabular}
\end{center}

\vspace{0.5cm}

%\newpage

   \begin{figure}[tb]\label{fig-1}
\unitlength=1mm
\vskip -1.5cm
\begin{picture}(0,100)
  \put(0,-5){%
   \psfig{file=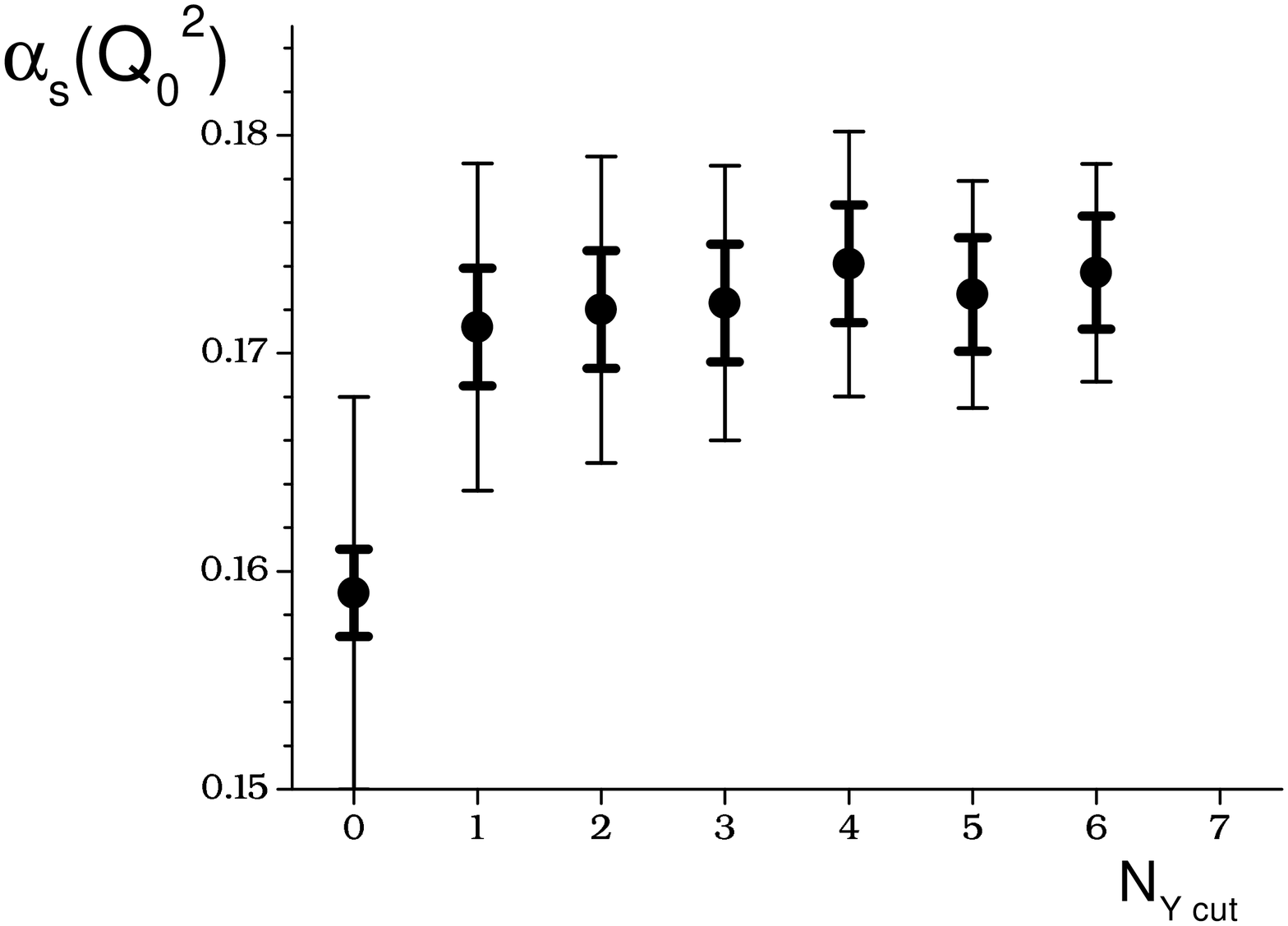,width=170mm,height=90mm}%
}
\end{picture}
\vskip 0cm
 \caption{
The study of systimatics at different $y_{cut}$ values
in the fits based on nonsinglet evolution.
The  QCD analysis of BCDMS $C^{12}, H_2, D_2$ data (nonsinglet case):
$x_{cut}=0.25$, $Q_0^2=90$ GeV$^2$. Thresholds of $c$ and $b$ quarks
are chosen at $Q^2=9$ GeV$^2$ and $Q^2=80$ GeV$^2$, respectively.  
The inner (outer) error-bars show statistical (systematic) errors.
}
\vskip 0cm
 \end{figure}
%\end{document}

From the Table 2 and 
the Fig. 1 we can see that the $\alpha_s$ values are obtained
for $N=1 \div 6$ of $Y_{cut3}$, $Y_{cut4}$ and $Y_{cut5}$ are very stable and
statistically consistent. The case $N=6$ reduces the systematic error
in $\alpha_s$ by factor $1.8$ and increases the value of $\alpha_s$,
while increasing the statistical error on the 30\%.

After the cuts have been implemented (in this Section below we use the set 
$N=6$),
we have 452 points in the analysis.
Fitting them in agreement with the same procedure considered in the previous 
Section,
we obtain the following results:
\bea
\as(90~\mbox{GeV}^2) &=& 0.1737 \pm 0.0029 ~\mbox{(stat)} 
\pm 0.0050 ~\mbox{(syst)} \pm 0.0025 ~\mbox{(norm)} 
\nonumber \\
& & \label{bd1.a} \\
\as(M_Z^2) &=& 0.1153 \pm 0.0013 ~\mbox{(stat)} 
\pm 0.0022 ~\mbox{(syst)} \pm 0.0012 ~\mbox{(norm)},
\nonumber
\eea
where
hereafter the symbol 
%``stat'', ``syst'' and 
``norm'' marks the 
%statistical error, systematic one and the 
error of normalization of experimental data.
%respectively.
Thus, the last error ($\pm 0.0011$ to $\as(M_Z^2)$) comes from difference 
in fits with free and fixed normalizations of BCDMS data 
\cite{BCDMS1,BCDMS2,BCDMS3} having 
different values of energy.\\

So, for the fits with NS evolution of BCDMS data 
\cite{BCDMS1,BCDMS2,BCDMS3} with minimization of
systematic errors, we have the following results:
\bea
\as(M_Z^2) ~=~ 0.1153 \pm 0.0028 ~\mbox{(total experimental error)} 
\label{bd1.a1}
\eea
Here 
total experimental error is squared root of sum of squares of 
statistical error, systematic one and error of normalization.

The value of $\as(M_Z^2)$ corresponds to the following value 
of QCD mass parameter:
\bea
\Lambda^{(5)}_{\MSbar} &=& 
\biggl( 181 \pm 32 ~\mbox{(total experimental error)} 
\biggr) \mbox{ MeV}, \nonumber \\
\Lambda^{(4)}_{\MSbar} &=& 
\biggl( 257 \pm 40 ~\mbox{(total experimental error)} 
\biggr) \mbox{ MeV}
\label{bd11.a}
\eea

\vskip 0.5cm

{\bf 4.1.2. The study of $N_{max}$-dependence. }\\

Following to \cite{Kri, Kri1}, we study the dependence of our results
on the $N_{max}$ value.
The full set of data is 452 points. The $Q^2$-evolution starts at
$Q^2_0$=90 GeV$^2$.

As it can be seen in the Table 3, our results are very stable, 
%at $N_{max} \geq 4$, 
that is in very good agreement with 
\cite{Kri}.\\

{\bf Table 3.} The values of $\asMZ$ at different values of $N_{max}$.
\vspace{0.2cm}
\begin{center}
%\footnotesize
\begin{tabular}{|l|c|c|c|c|}
%\begin{tabular}{|p{35pt}|p{38pt}|p{38pt}|p{38pt}|p{38pt}|p{38pt}|p{38pt}|}
\hline
& & & &  \\
$N_{max}$  & $\chi^2(F_2)/DOF$ &  $\chi^2(\mbox{slope})$ 
& $\as(90~\mbox{GeV}^2)$  $\pm$ stat & 
$\asMZ$  $\pm$ stat \\
& &for $6$ points  &stat $=0.0038$  &stat $=0.0013$ \\
\hline \hline
3 & 1.08 & 7.3 & 0.1720 &  0.1155 \\  
%\hline
4 & 0.97 & 11.3 & 0.1715 & 0.1143  \\  
%\hline
5 & 1.11 & 6.9 & 0.1729 & 0.1144 \\  
%\hline
6 & 0.95 & 3.6 & 0.1747  & 0.1157 \\  
%\hline
7 & 0.94 & 5.4 & 0.1740 & 0.1154 \\  
%\hline
8 & 0.94 & 6.8 & 0.1738 & 0.1153 \\  
%\hline
9 & 0.94 & 7.6 & 0.1735 & 0.1152  \\  
%\hline
10 & 1.07 & 7.7 & 0.1735 & 0.1152 \\  
%\hline
11 & 1.08 & 7.2 & 0.1726 & 0.1149 \\  
%\hline
12 & 1.04 & 7.1 & 0.1731 & 0.1152 \\  
%\hline
13 & 1.11 & 7.1 & 0.1725 & 0.1149 \\  
\hline
\end{tabular}
\end{center}

\vspace{0.5cm}

Starting with $N_{max} =5$, where our results are already very stable, 
we put
the results together and can calculate average value of $\as(M_Z^2)=0.1152$
and estimate average deflection. The deflection is $0.0002$ and can be
considered as error of the Jacobi Polynomial expansion method, 
i.e. {\it method error}. 

\subsection { SLAC and NMC $H_2 + D_2$ data and BFP %iron 
$Fe$ data}

We continue our NS evolution
analyses by fits of
%with 
experimental data \cite{SLAC1,SLAC2,NMC,BFP}
obtained 
%many years ago 
by SLAC, NM and BFP Collaborations.
The full set of data is 345 points (when $x \geq 0.25$): 
238 ones of SLAC, 66 ones of NMC and 41 ones of BFP.
The starting point of QCD evolution is $Q^2_0=20$ GeV$^2$, 
%HT terms are included, 
the $Q^2$-cut is $Q^2>1$ GeV$^2$. \\

For illustration of importance of $1/Q^2$ corrections
we fit the data in the following way. First of all, we analyze the data 
applying only perturbative QCD part of SF $F_2$, i.e. $F_2^{tw2}$. Later, 
we have added $1/Q^2$ corrections: firstly, target mass ones and later
twist-four ones. As it is possible to see in the Table 4, we have the very 
bad fit, when we work only with twist-two part $F_2^{tw2}$.
The agreement with the data is improved essentially when target mass 
corrections have been added. The incorporation of twist-four corrections 
leads to very good fit of the 
%SLAC 
data. 
Neglect of systematic errors deteriorates twice our results.
We combine the statistical and systematic errors in quadrature.\\

%\hskip -.56cm
%
{\bf Table 4.} The values of $\asMZ$ and $\chi^2$ at different 
regimes of fits.
\vspace{0.2cm}
\begin{center}
%\footnotesize
\begin{tabular}{|l|c|c|c|c|c|c|c|c|}
\hline
& & & & & & & \\
$N ~~~of$  & TMC & HTC & syst. &$\chi^2(F_2)/DOF$ & $\chi^2(slope)$ 
& $\as(20~\mbox{GeV}^2)$ &
$\asMZ$ \\
fits & & & error &  &for $6$ points  &$\pm$ stat  &   \\
\hline \hline
1 & No & No &  Yes & 6.0 & 1050 & 0.2131 $\pm$ 0.0012 & 
0.1167 \\  
%\hline
2 & Yes & No &  Yes & 2.3 & 224 & 0.2017 $\pm$ 0.0013 & 
0.1133 \\  
%\hline
3 & Yes & Yes & No & 1.8 & 12.0 & 0.2230 $\pm$ 0.0030 & 
0.1195 \\  
%\hline \hline
4 & Yes & Yes & Yes & 0.8 & 6.1 & 0.2231 $\pm$ 0.0060 & 
0.1195 \\  
\hline
\end{tabular}
\end{center}

\vspace{0.5cm}

We have got the following values for parameters in parameterizations of
parton distributions (at $Q^2_0=20$ GeV$^2$):
\bea
A_{NS}^P &=& 1.44,~~~~~~~~ A_{NS}^D ~=~ 2.06,~~~~~~~~~ A_{NS}^F ~=~ 1.87,
  \nonumber \\
b_{NS}^P &=& 3.88,~~~~~~~~~ b_{NS}^D ~=~ 3.84,~~~~~~~~~ b_{NS}^F ~=~ 4.23,
  \nonumber \\
d_{NS}^P &=& 10.9,~~~~~~~~~ d_{NS}^D ~=~ 4.04,~~~~~~~~~ d_{NS}^F ~=~ 5.03,  
\label{para}
\eea
where
the symbols $P$, $D$ and $F$ denote the parameters of parameterizations
for proton, deuteron and iron data, respectively.

We note that the
%The 
values of the coefficients are close to ones obtained in other 
numerical analyses
(see \cite{Al2000,Al2001,KPS,KPS1} and references therein).
%Indeed, 
The values of $b_{NS}^l$ ($l=P,D,F$) are in quite good agreement with 
quark-counting rules of Ref.\cite{schot}. There is also good agreement
with 
%fits \cite{Al2000,Al2001,KPS,KPS1} and with 
theoretical studies
\cite{Gross,VoKoMa}.\\

{\bf Table 5.} The values of the twist-four terms.
%$\asMZ$ and $\chi^2$ at different regimes of fits.
\vspace{0.2cm}
\begin{center}
%\footnotesize
\begin{tabular}{|l||c|c|}
\hline
& &   \\
$x_i$ & $\tilde h_4(x_i)$ of $H_2$ &  $\tilde h_4(x_i)$ of $D_2$  \\
 & $\pm$ stat &  $\pm$ stat    \\
\hline \hline
0.25 & -0.149 $\pm$ 0.015  &  -0.176 $\pm$ 0.014 \\  
0.35 & -0.151 $\pm$ 0.013  &  -0.178 $\pm$ 0.012 \\  
0.45 & -0.214 $\pm$ 0.012  &  -0.147 $\pm$ 0.022 \\  
0.55 & -0.228 $\pm$ 0.022  &  -0.065 $\pm$ 0.037 \\  
0.65 &  0.024 $\pm$ 0.070  &   0.053 $\pm$ 0.080 \\  
0.75 &  0.227 $\pm$ 0.154  &   0.130 $\pm$ 0.131 \\  
\hline
\end{tabular}
\end{center}

\vspace{0.5cm}

The values of parameters of twist-four term are given in the Table 5.
We would like to note that the twist-four terms for $H_2$ and $D_2$ data
coincide with each other with errors taken into account. It is in
full agreement with analogous analysis \cite{ViMi}.

We obtain the following results (at $\chi^2(F_2)=250$, $\chi^2(slope)=6.1$
on $7$ points):
\bea
\as(20~\mbox{GeV}^2) &=& 0.2231 \pm 0.0060 ~\mbox{(stat)} 
\pm 0.0075 ~\mbox{(syst)} 
+ 0.0030~\mbox{(norm)}
\nonumber \\
& & \label{slo1} \\
%\nonumber \\
\as(M_Z^2) &=& 0.1195 \pm 0.0017 ~\mbox{(stat)} 
%+ \biggl\{ \begin{array}{l}+ 0.0020 \\ - 0.0034 \end{array} 
\pm 0.0022~\mbox{(syst)} 
+ 0.0010\mbox{(norm)} \nonumber
%\label{slo1}
\eea
The last error ($\pm 0.0010$ to $\as(M_Z^2)$) comes from fits 
with free and fixed normalizations between  different data of SLAC,
NM and BFP Collaborations.\\

So, the fits of SLAC, NMC and BFP data based on the nonsinglet evolution
give 
%the following result 
for coupling constant:
\bea
\as(M_Z^2) ~=~ 0.1195 \pm 0.0030 ~\mbox{(total experimental error)}, 
\label{bd1.2}
\eea
which corresponds to the following value 
of QCD mass parameter:
\bea
\Lambda^{(5)}_{\MSbar} &=& 
\biggl( 231 \pm 37 ~\mbox{(total experimental error)} 
\biggr) \mbox{ MeV}, \nonumber \\
\Lambda^{(4)}_{\MSbar} &=& 
\biggl( 321 \pm 44 ~\mbox{(total experimental error)} 
\biggr) \mbox{ MeV},
\label{bd11.1}
\eea
where the error connected with the type of normalization of data
are included already to systematic error.\\

Looking at
%together 
the results obtained in two previous subsections
we see good agreement (within existing errors) between the values of
the coupling 
constant $\as(M_Z^2)$ obtained
separately in the fits of BCDMS data and ones in the fits of combine 
SLAC, NMC and BFP data (see Eqs. (\ref{bd1.a})-(\ref{bd11.a}) and 
(\ref{slo1})-(\ref{bd11.1})).
Thus, we have possibility to fit together all the data that
is the subject of the following subsection.

\subsection { SLAC, BCDMS, NMC and BFP data }
%{\bf 4.1.} {\bf SLAC data}

 We use the following common $x$-cut: $x \geq
0.25 $ and $Y_{cut}$ with $N=6$ (see the Table 1) for the BCDMS data.
After these cuts have been incorporated, the full set of data is 797 points.
The starting point of QCD evolution is $Q^2_0=90$ GeV$^2$.\\

{\bf 4.3.1. The results of fits. }\\

We verify here the range of applicability of perturbative QCD. To do it,
we analyze firstly the data without a contribution of twist-four terms,
i.e. when $F_2 = F_2^{pQCD}$. We do several fits using the cut 
$Q^2 \geq Q^2_{cut}$ and increase the value $Q^2_{cut}$ step by step.
We observe  good agreement of the fits with the data when 
$Q^2_{cut} \geq 10$ GeV$^2$ (see the Table 6).

Later we add the twist-four corrections and fit the data with the
usual cut $Q^2 \geq 1$ GeV$^2$.
We have find very good agreement with the data. Moreover 
the predictions for $\asMZ$ in both above procedures 
are very similar (see the Table 6 and Fig. 2).\\

{\bf Table 6.} The values of $\asMZ$ and $\chi^2$ at different 
regimes of fits.

\vspace{0.2cm}
%\begin{center}
\begin{tabular}{|l|c|c|c|c|c|c|}
\hline
& & &  & &  &\\
$N$ of & $Q^2$ & $N$ of & HTC &$\chi^2(F_2)$/DOF &  
$\as(90~\mbox{GeV}^2)$ $\pm$ stat & $\asMZ$ \\
fits & cut & points &  &  & &  \\
\hline \hline
1 & 1.0 & 797 &  No & 2.87 & 0.1679 $\pm$ 0.0007 & 0.1128  \\  
%\hline
2 & 2.0 & 772 &  No & 1.82 & 0.1733 $\pm$ 0.0007 & 0.1151 \\  
%\hline
3 & 3.0 & 745 &  No & 1.38 & 0.1789 $\pm$ 0.0009 &  0.1175 \\  
%\hline
4 & 4.0 & 723 &  No & 1.23 & 0.1802 $\pm$ 0.0009 & 0.1180 \\  
%\hline
5 & 5.0 & 703 &  No & 1.19 & 0.1813 $\pm$ 0.0011 & 0.1185 \\  
%\hline
6 & 6.0 & 677 &  No & 1.13 & 0.1803 $\pm$ 0.0013 & 0.1189 \\  
%\hline
7 & 7.0 & 650 &  No & 1.09 & 0.1799 $\pm$ 0.0016 & 0.1179  \\  
%\hline
8 & 8.0 & 632 &  No & 1.06 & 0.1803 $\pm$ 0.0019 & 0.1181 \\  
%\hline
9 & 9.0 & 613 &  No & 1.01 & 0.1797 $\pm$ 0.0023 & 0.1178 \\  
%\hline
10 & 10.0 & 602 &  No & 0.98 & 0.1776 $\pm$ 0.0022 & 0.1170 \\  
11 & 11.0 & 688 &  No & 0.97 & 0.1770 $\pm$ 0.0024 & 0.1167 \\  
12 & 12.0 & 574 &  No & 0.97 & 0.1768 $\pm$ 0.0025 & 0.1167 \\  
\hline \hline
13 & 1.0 & 797 & Yes & 0.97 & 0.1785 $\pm$ 0.0025 & 0.1174  \\  
\hline
\end{tabular}

\vspace{0.5cm}

We have got the following values for parameters in parameterizations of
parton distributions (at $Q^2_0=90$ GeV$^2$):

\bea
A^P_{NS} &=& 2.40,~~\, A^D_{NS} ~=~ 2.46,~~~ A^C_{NS} ~=~ 2.46,
~~~ A^{F}_{NS} ~=~ 1.65,
  \nonumber \\
b^P_{NS} &=& 3.98,~~~\, b^D_{NS} ~=~ 3.94,~~~\,\, b^C_{NS} ~=~ 4.08,~~~\,\, 
b^{F}_{NS} ~=~ 4.72,
  \nonumber \\
d^P_{NS} &=& 4.85,~~~\, d^D_{NS} ~=~ 2.38,~~~\, d^C_{NS} ~=~ 1.55,~~~\,\, 
d^{F}_{NS} ~=~ 7.97,
\label{para1}
\eea

The values 
%of parameters 
are in good agreement with ones presented
in previous subsection. Then all discussions given there can
be applied here.\\

{\bf Table 7.} The values of the twist-four terms.
\vspace{0.2cm}
\begin{center}
\begin{tabular}{|l||c|c||l||c|}
\hline
& & & &  \\
$x_i$ & $\tilde h_4(x_i)$ of $H_2$ &  $\tilde h_4(x_i)$ of $D_2$ &
$x_i$ & $\tilde h_4(x_i)$ of $C$ and $Fe$  \\
 & $\pm$ stat & $\pm$ stat  & &  $\pm$ stat   \\
\hline \hline
0.275 & -0.221 $\pm$ 0.010  &  -0.226 $\pm$ 0.010 &  
0.250 & -0.118 $\pm$ 0.187 \\
0.350 & -0.252 $\pm$ 0.010  &  -0.214 $\pm$ 0.010 &  
0.350 & -0.415 $\pm$ 0.233 \\
0.450 & -0.232 $\pm$ 0.019  &  -0.159 $\pm$ 0.020 &  
0.450 & -0.656 $\pm$ 0.494 \\
0.550 & -0.122 $\pm$ 0.360  &  -0.058 $\pm$ 0.300 &  
& \\
0.650 & -0.159 $\pm$ 0.031  &  -0.057 $\pm$ 0.031 &  
 & \\
0.750 &  0.040 $\pm$ 0.050  &   0.020 $\pm$ 0.049 &  
 &  \\
\hline
\end{tabular}
\end{center}

\vspace{0.5cm}

%\newpage

\begin{figure}[tb]
\vskip -0.5cm
\begin{center}
\psfig{figure=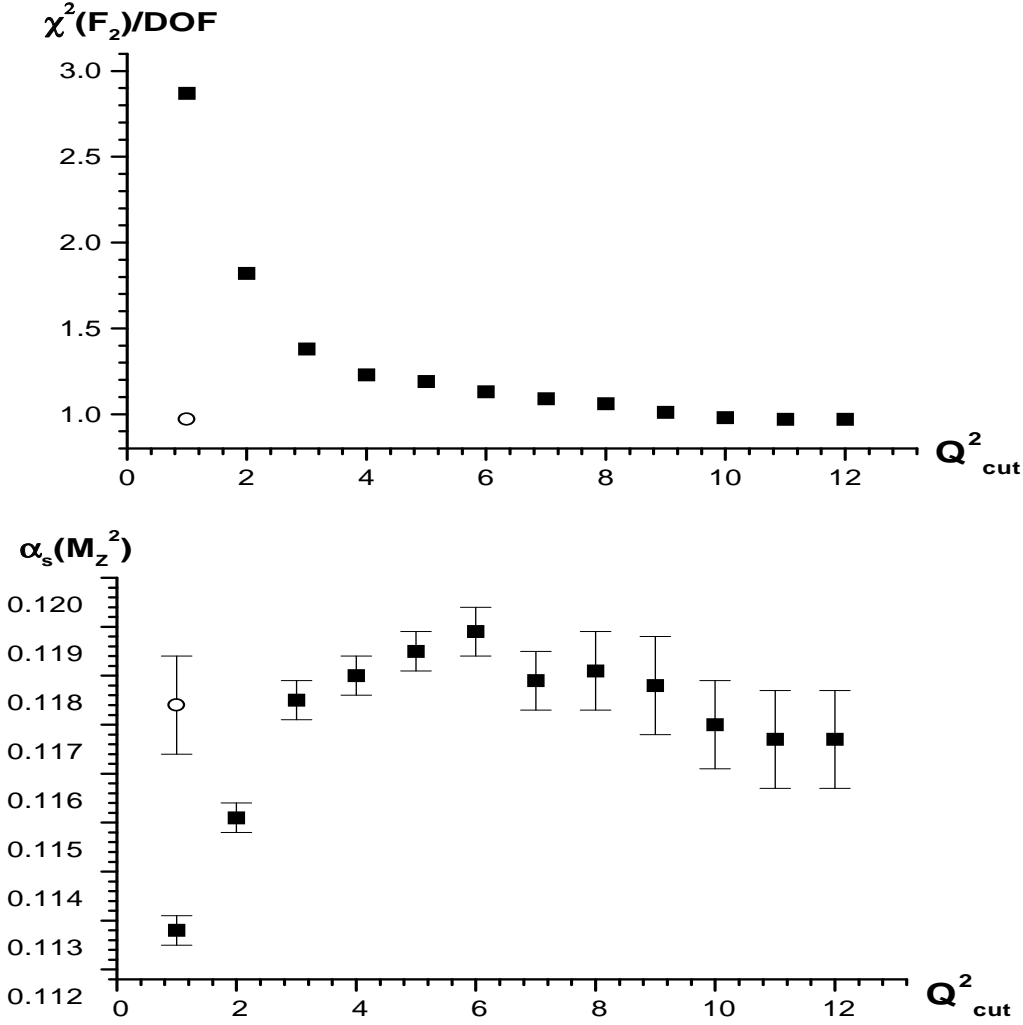,width=14cm,height=15cm}
\end{center}
\vskip -1cm
\caption{
The values of $\asMZ$ and $\chi^2$ at different $Q^2$-values of data cutes
in the fits based on nonsinglet evolution.
%regimes of fits. 
The black 
%and white 
points show the 
analyses of data without  twist-four contributions.
The white point corresponds to the case where twist-four contributions 
were added.  Only statistical errors are shown.
}
\label{Non}
\end{figure}

The Table 7 contains the value of parameters of the twist-four term.
As it was in the previous subsection,
%We would like to note that 
the twist-four terms for $H_2$ and $D_2$ data
coincide with each other with errors taken into account that is in 
agreement with \cite{ViMi}.\\

%\vspace{0.5cm}

So, the analysis of combine SLAC, NMC, BCDMS and BFP data 
are given the following results:
\begin{itemize}
\item  
When HT corrections are not included and the cut of $Q^2$ is 10 GeV$^2$
at the free normalization 
\bea 
\chi^2\mbox{/DOF}~=~0.98~~~\mbox{ and }~~~
\as(90~\mbox{GeV}^2) &=& 0.1776 \pm 0.0022 ~\mbox{(stat)}, \nonumber \\
& &  \nonumber \\
\as(M_Z^2) &=& 0.1170 \pm 0.0009 ~\mbox{(stat)}
\label{fu1.1}
\eea
\item  
When HT corrections are included and the cut of $Q^2$ is 1 GeV$^2$ 
\bea 
\chi^2\mbox{/DOF}~=~0.97~~~\mbox{ and }~~~
\as(90~\mbox{GeV}^2) &=& 0.1785 \pm 0.0025 ~\mbox{(stat)}, \nonumber \\
& & \nonumber \\
\as(M_Z^2) &=& 0.1174 \pm 0.0010 ~\mbox{(stat)}
\label{fu1.2}
\eea
\end{itemize} 

Thus, as it follows from nonsinglet fits of experimental data, 
perturbative QCD
works rather well at $Q^2 \geq 10$ GeV$^2$.\\

{\bf 4.3.2. The study of threshold effects. }\\

Here we would like to study threshold effects in $Q^2$-evolution of SF $F_2$.
Note that at NLO level in nonsinglet case the coefficient function of $F_2$
and anomalous dimension do not depend on the number $n_f$ of active quarks.
Then, the our study of the threshold effects in $Q^2$-evolution of SF 
$F_2$ is exactly equal to the investigation of a role of threshold effects 
in the QCD coupling constant $\as(Q^2)$.

To study the threshold effects we consider two types of possible thresholds
of heavy quarks: $Q^2_f = 4m^2_f$ and $Q^2_f = m^2_f$. First type of
thresholds has appeared when a heavy quark with the mass $m_f$ takes a
possibility to be born. The second one lies close to the position
of ``Euclidean-reflected'' threshold of heavy quarks. It should play
a significant role (see \cite{SSiMi1}) in the $\as(Q^2)$-evolution.\\

{\bf A.~~ } Let thresholds appear
%take places 
at $Q^2_f=4m_f^2$. 
Then we split the range
of the data to three separate ones: 
\begin{itemize}
\item  The $Q^2$ values are between $1$ GeV$^2$ and $10$ GeV$^2$, where
the number $n_f$ of active quarks is $3$.
\item  The $Q^2$ values are between $10$ GeV$^2$ and $80$ GeV$^2$, where
the number $n_f$ of active quarks is $4$.
\item  The $Q^2$ values are above $80$ GeV$^2$, where
the number $n_f$ of active quarks is $5$.
\end{itemize} 

\vspace{0.5cm}

{\bf Table 8.} The values of $\asMZ$ and $\chi^2$ at different 
regimes of fits.
\vspace{0.2cm}
\begin{center}
%\footnotesize
\begin{tabular}{|l|c|c|c|c|c|c|c|c|c|}
\hline
& & & & & & & & &\\
$N$ of & $Q^2$ & $n_f$ & $Q^2_0$ & $N$ of &$\chi^2(F_2)$&  
$\Lambda_{\overline{MS}}^{(3)}$ & 
$\Lambda_{\overline{MS}}^{(4)}$ &  $\Lambda_{\overline{MS}}^{(5)}$ &
$\asMZ$ \\
fit & range &  &  & points &  & $\pm$ stat & $\pm$ stat & $\pm$ stat 
& $\pm$ stat \\
& &  &  &  &  & (MeV) & (MeV) & (MeV) &  \\
\hline \hline
1 & 1-10 & 3 & 5 & 195 & 124 & 400 $\pm$ 30 & 308 $\pm$ 26 & 220 $\pm$ 23 &
0.1187 $\pm$ 0.0020 \\  
%\hline
2 & 10-80 & 4 & 20 & 455 & 471 & & 291 $\pm$ 17 & 208 $\pm$ 13 &
0.1177 $\pm$ 0.0012 \\  
%\hline
3 & 80-300 & 5 & 90 &190 & 143 & & & 199 $\pm$ 54 &
0.1169 $\pm$ 0.0040 \\  
\hline
\end{tabular}
\end{center}

\vspace{0.5cm}

The results are shown in Table 8.
The average $\asMZ$ value can be calculated 
%from above Table. It 
and it has the following
value:
\bea 
\as(M_Z^2) &=& 0.1178 \pm 0.0010 ~\mbox{(stat)}
\label{fu1.3}
\eea

\vspace{0.5cm}

{\bf B.~~ } Let thresholds appear
%take places 
at $Q^2_f=m_f^2$. 
Then we split the range of the data to two separate ones: 
\begin{itemize}
\item  The $Q^2$ values are between $2.5$ GeV$^2$ and $20.5$ GeV$^2$, where
the number $n_f$ of active quarks is $4$.
\item  The $Q^2$ values are above $20.5$ GeV$^2$, where
the number $n_f$ of active quarks is $5$.
\end{itemize} 

\vspace{0.5cm}

%\hskip -.56cm
%
{\bf Table 9.} The values of $\asMZ$ and $\chi^2$ at different 
regimes of fits.
\vspace{0.2cm}
\begin{center}
%\footnotesize
\begin{tabular}{|l|c|c|c|c|c|c|c|c|}
\hline
& & & & & & & &\\
$N$ of & $Q^2$ & $n_f$ & $Q^2_0$ & $N$ of &$\chi^2(F_2)$&   
$\Lambda_{\overline{MS}}^{(4)}$ &  $\Lambda_{\overline{MS}}^{(5)}$ &
$\asMZ$ \\
fit & range &  &  & points &  & $\pm$ stat & $\pm$ stat 
& $\pm$ stat \\
& &  &  &  &  & (MeV) & (MeV) &  \\
\hline \hline
1 & 2.5-20.5 & 4 & 10 & 241 & 197 & 298 $\pm$ 10 & 213 $\pm$ 8 &
0.1181 $\pm$ 0.0007 \\  
%\hline
2 & 20.5-300 & 5 & 90 & 558 & 533 & & 187 $\pm$ 16 &
0.1159 $\pm$ 0.0014 \\  
\hline
\end{tabular}
\end{center}

\vspace{0.5cm}

The results are shown in Table 9.
The average $\asMZ$ value can be calculated 
%from above Table. It 
and it has the following
value:
\bea 
\as(M_Z^2) &=& 0.1176 \pm 0.0006 ~\mbox{(stat)}
\label{fu1.4}
\eea

Thus, we do not find a strong dependence on exact value of thresholds
of heavy quarks.
The theoretical uncertainties due to threshold effects can be estimated for
$\as(M_Z^2)$ as $0.0002$.

\subsection {The results of the analyses based on nonsinglet evolution }
%{\bf 4.3.3. The results of NS analyses }\\

Thus, using the analyses based on NS evolution of the SLAC, NMC, BCDMS 
and BFP experimental data for SF $F_2$ we obtain for 
$\asMZ$ the following expressions:\\

{\bf 1.} When we switch off the twist-four corrections, and put the cut
%no HTC, 
$Q^2 > 10$ GeV$^2$, we have got
% and  
at $\chi^2/DOF=0.98$
\bea
\as(M_Z^2) ~=~ 0.1170 \pm 0.0009 ~\mbox{(stat)} 
\pm 0.0019 ~\mbox{(syst)}  \pm 0.0010 ~\mbox{(norm)}
\label{NSfin}
\eea
or
\bea
\as(M_Z^2) ~=~ 0.1170 \pm 0.0023 ~\mbox{(total experimental error)} 
\label{NSfin1}
\eea

\vspace{0.5cm}

{\bf 2.} When we add the twist-four corrections, and put the cut
$Q^2 > 1$ GeV$^2$, we have got
%and  
at $\chi^2/DOF=0.97$
\bea
\as(M_Z^2) ~=~ 0.1174 \pm 0.0007 ~\mbox{(stat)} 
\pm 0.0021 ~\mbox{(syst)}  \pm 0.0005 ~\mbox{(norm)} 
\label{NSfin2}
\eea
or
\bea
\as(M_Z^2) ~=~ 0.1174 \pm 0.0022 ~\mbox{(total experimental error)} 
\label{NSfin3}
\eea

\vspace{0.5cm}

Looking at the results obtained in the Section
we see that the central value of the coupling constant $\as(M_Z^2)$ obtained
in the fits (based on NS evolution) 
of combine SLAC, BCDMS, NM and BFP data lies between
the central values of the coupling constants
%$\as(M_Z^2)$ 
obtained separately in the fits of 
BCDMS data and in ones of SLAC, BCDMS, NM and BFP data.
All obtained values of $\as(M_Z^2)$
%the coupling constants 
are in good agreement within existing statistical
%experimental 
errors.

\begin{figure}[tb]
\vskip 0cm
\begin{center}
\psfig{figure=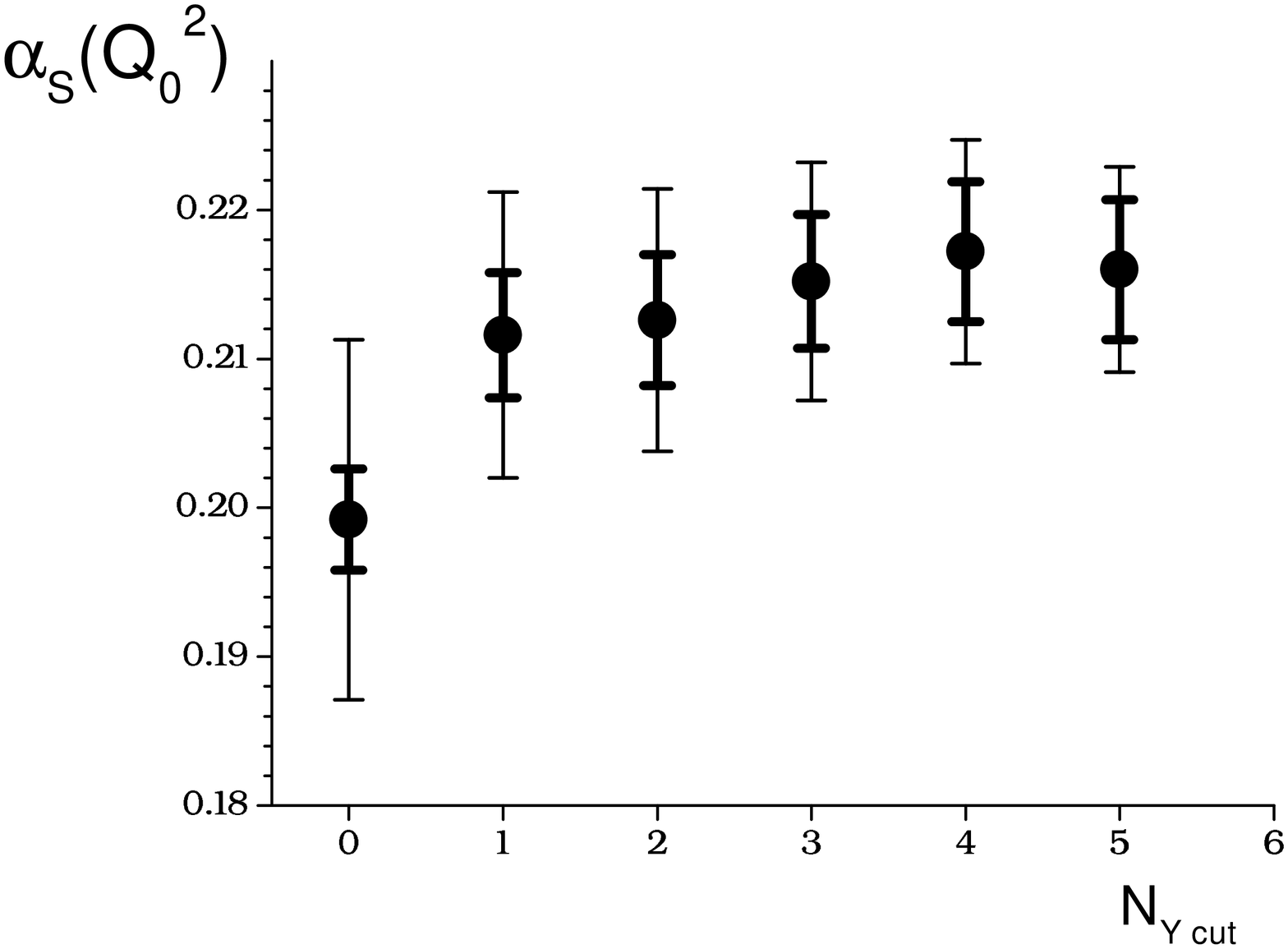,width=17cm,height=9cm}
\end{center}
\vskip 0cm
 \caption{
The study of systimatics at different $y_{cut}$ values
in the fits based on combine siglet and nonsinglet evolution. 
The  QCD analysis of BCDMS $C^{12}, H_2, D_2$ data (the case of
combine evolution):
%(singlet case):
no a $x_{cut}$, $Q_0^2=20$ GeV$^2$. Thresholds of $c$ and $b$ quarks
are chosen at $Q^2=9$ GeV$^2$ and $Q^2=80$ GeV$^2$, respectively.  
The inner (outer) error-bars show statistical (systematic) errors.
}
\vskip 1cm
 \end{figure}

\section{  Results of fits of $F_2$: the combined nonsinglet and \\
singlet evolution
}

%\subsection { SLAC data (ne budem !!!)}

At this case, the quite low $x$ experimental data lie at low
$Q^2$ range and we choose $Q^2_0$ = 20 GeV$^2$.
We
%For the singlet case we 
use also
$N_{max} =8$.

The study of the $N_{max}$-dependence of the results in the
combine nonsinglet and singlet case of evolution
has been found in \cite{Kri1}. Note here only that the analysis in
\cite{Kri1} shows the $N_{max}$-independence of the obtained results
starting already with $N_{max} =7$.

\subsection { BCDMS $C^{12} + H_2 + D_2$ data }
%{\bf 4.1.} {\bf SLAC data}

As in the previous Section, we start our analyses with the 
%most precise 
experimental data 
\cite{BCDMS1,BCDMS2,BCDMS3} obtained  by BCDMS muon
%$\mu h$ 
scattering experiment.
The full set of data is 762 points.
% (when $x \geq 0.25$).
The starting point of QCD evolution is $Q^2_0=20$ GeV$^2$.\\

\begin{figure}[tb]
\vskip 0cm
\begin{center}
\epsfig{figure=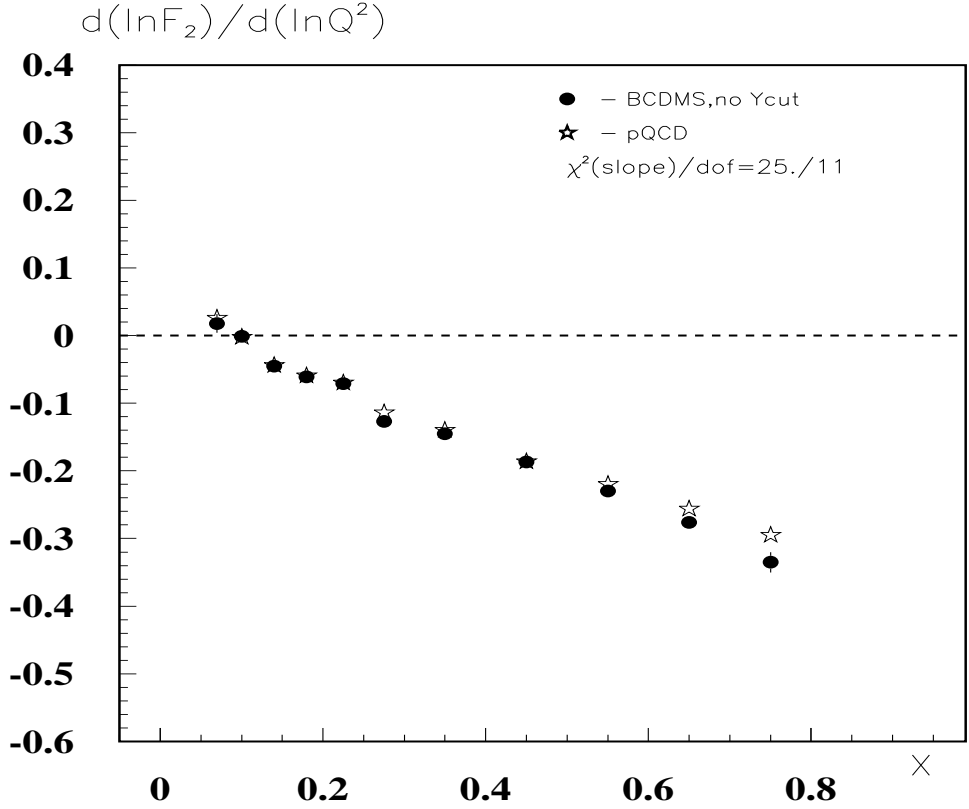,width=15cm,height=12cm}
\end{center}
\vskip -1.5cm
 \caption{
The values of the slope $d(\ln{F_2})/d(\ln{Q^2})$ at $Q^2=20$ GeV$^2$.
The white points correspond to the theoretical predictions
based on combine singlet and nonsinglet evolution.
% and $Q^2_0=20$ GeV$^2$, respectively.
The black points show BCDMS $C^{12}$, $H_2$ and $D_2$ data without
%no a $x_{cut}$ and 
a $Y_{cut}$. 
}
%\vskip 1cm
 \end{figure}

\begin{figure}[tb]
\vskip 0cm
\begin{center}
\epsfig{figure=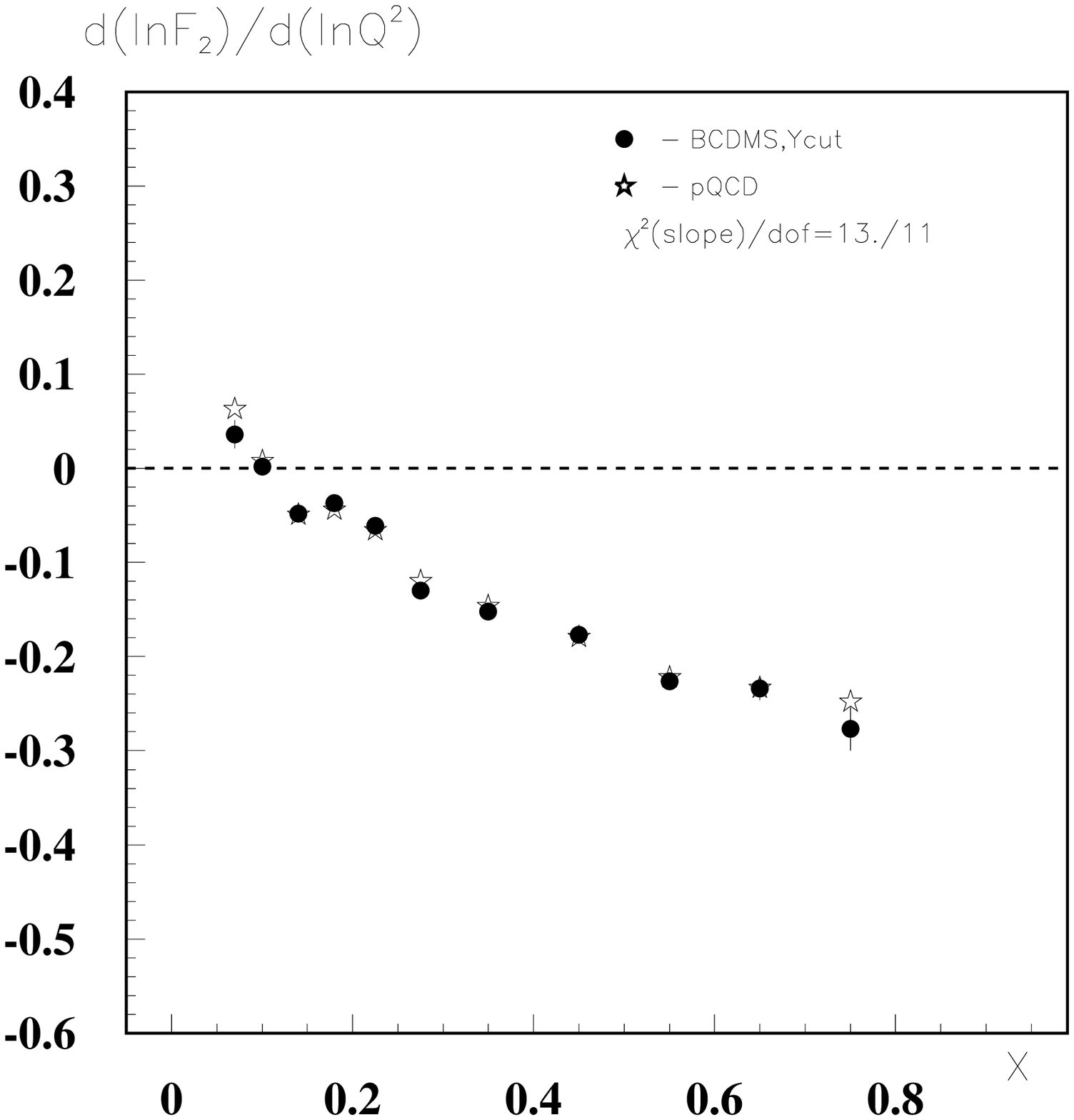,width=15cm,height=12cm}
\end{center}
\vskip -1.5cm
 \caption{
Notation as in Fig. 4 with one exception: the $Y_{cut}$ with
$N=5$ is taken into account.
}
%\vskip 1cm
 \end{figure}

As in the nonsinglet evolution case we have studied influence of the 
experimental systematic errors on the
results of the QCD analysis as a function of $Y_{cut3}$, $Y_{cut4}$ and
$Y_{cut5}$ applied to the data. Here
we use also
several sets $N$ of the values for the cuts at $0.5 < x \leq 0.8$, 
which are given in the Table 10.

The absolute differences between the values of $\alpha_s$ for the distorted
and undistorted sets of data are given in Table 11 and the Fig. 3
as the total systematic
error of $\alpha_s$ estimated in quadratures. The number of the experimental
points and the value of $\alpha_s$ for the undistorted set of $F_2$ are also
given in the Table 11 and the Fig. 3. \\

%\newpage
%\hskip -.56cm
%
{\bf Table 10.} The values of $Y_{cut3}$, $Y_{cut4}$ and $Y_{cut5}$.
\vspace{0.2cm}
\begin{center}
%\footnotesize
\begin{tabular}{|c|c|c|c|c|c|c|}
\hline
& & & & & &  \\
$N$ & 0 & 1 & 2 & 3 & 4 & 5  \\
& & & & & &  \\
\hline \hline
$Y_{cut3}$ & 0 & 0.14 & 0.16 & 0.18 & 0.22 & 0.23 \\  
%\hline
$Y_{cut4}$ & 0 & 0.16 & 0.18 & 0.20 & 0.23 & 0.24 \\
$Y_{cut5}$ & 0 & 0.20 & 0.20 & 0.22 & 0.24 & 0.25 \\
\hline
\end{tabular}
\end{center}

\vspace{0.5cm}

%\newpage
{\bf Table 11.} The values of $\asMZ$ at different values of $N$.
\vspace{0.2cm}
\begin{center}
%\footnotesize
\begin{tabular}{|l|c|c|c|c|}
%\begin{tabular}{|p{35pt}|p{38pt}|p{38pt}|p{38pt}|p{38pt}|p{38pt}|p{38pt}|}
\hline
& & & &  \\
$N$  & number & $\chi^2(F_2)/DOF$  & $\as(20~\mbox{GeV}^2)$  $\pm$ stat 
&full  \\
&of points &  & &syst. error \\
\hline \hline
0 & 762 & 1.22 &  0.1992 $\pm$ 0.0034 & 0.0122 \\
1 & 649 & 1.06 &  0.2116 $\pm$ 0.0042 & 0.0096 \\
2 & 640 & 1.07 &  0.2126 $\pm$ 0.0044 & 0.0088 \\
3 & 627 & 1.05 &  0.2152 $\pm$ 0.0045 & 0.0080 \\
4 & 596 & 1.04 &  0.2172 $\pm$ 0.0047 & 0.0076 \\
5 & 590 & 1.04 &  0.2160 $\pm$ 0.0047 & 0.0068 \\
\hline
\end{tabular}
\end{center}

\vspace{0.5cm}

From the Table 11 and the Fig. 3 we can see that the $\alpha_s$ values are 
obtained
for $N=1 \div 5$ of $Y_{cut3}$, $Y_{cut4}$ and $Y_{cut5}$ are very stable and
statistically consistent. The case $N=5$ reduces the systematic error
in $\alpha_s$ by factor $1.8$ and increases the value of $\alpha_s$,
while increasing the statistical error on the 27\%.

The importance of the $Y$-cut can be shown also in the Figs. 4 and 5, where
the slope \\ $d(\ln{F_2})/d(\ln{Q^2})$ has been shown 
at 
%$Q^2=??$ GeV$^2$ and 
$Q^2=20$ GeV$^2$. As we can see, there
is an essential inprovement (the corresponding $\chi^2$(slope) decreases
in half), when the $Y$-cut has been taken into account.\\

\begin{figure}[tb]
\vskip 0cm
\begin{center}
\epsfig{figure=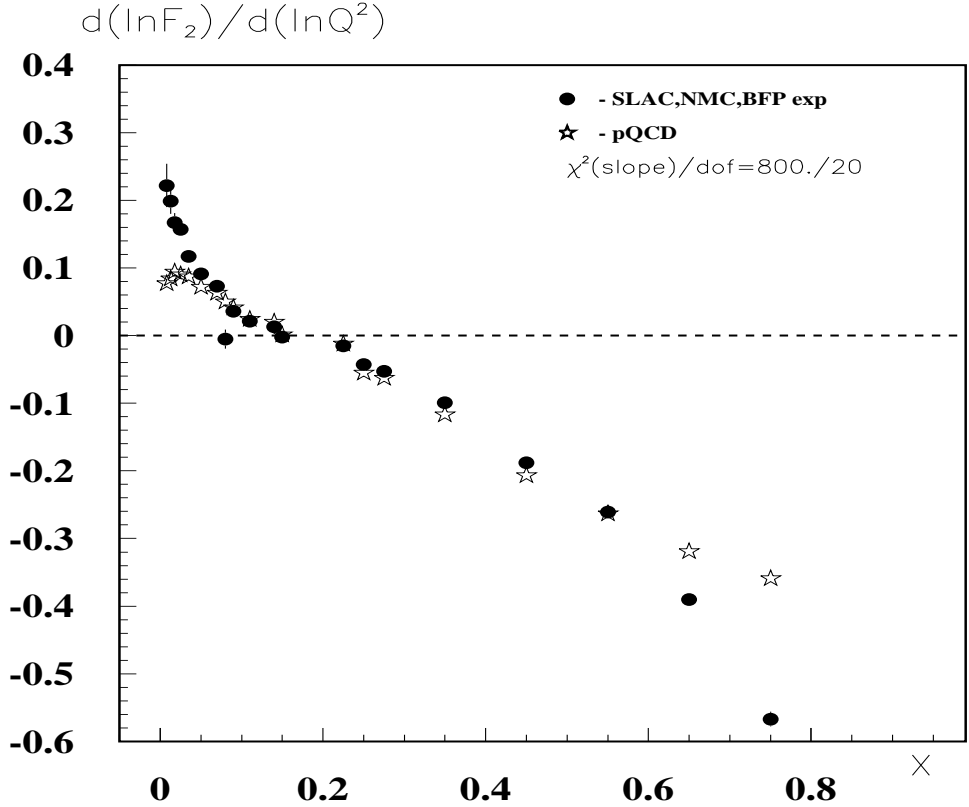,width=15cm,height=12cm}
\end{center}
\vskip -1.5cm
 \caption{
The values of the slope $d(\ln{F_2})/d(\ln{Q^2})$ at $Q^2=20$ GeV$^2$.
The white points correspond to the theoretical predictions
based on the twist-two approximation of perturbative QCD and
combine singlet and nonsinglet evolution.
%at $Q^2=??$ GeV$^2$ and $Q^2_0=20$ GeV$^2$, respectively.
The black points show SLAC, NMC and BFP experimental data.
}
%\vskip 1cm
 \end{figure}

\begin{figure}[tb]
\vskip 0cm
\begin{center}
\epsfig{figure=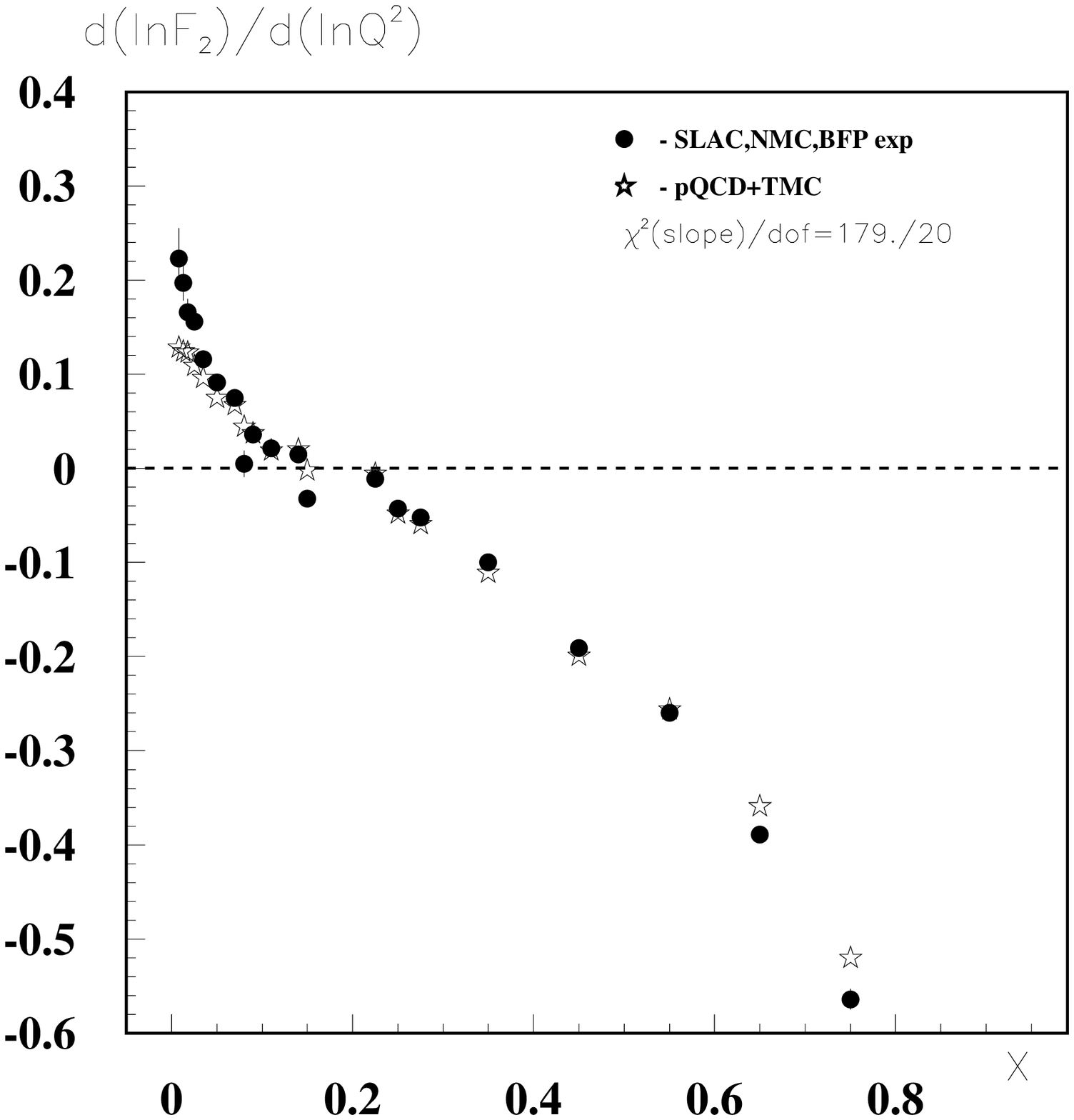,width=15cm,height=12cm}
\end{center}
\vskip -1.5cm
 \caption{
Notation as in Fig. 6 with one exception: the target mass corrections
%$y_{cut}$ with$N=5$ is 
are taken into account for theoretical predictions.
}
%\vskip 1cm
 \end{figure}

\begin{figure}[tb]
\vskip 0cm
\begin{center}
\epsfig{figure=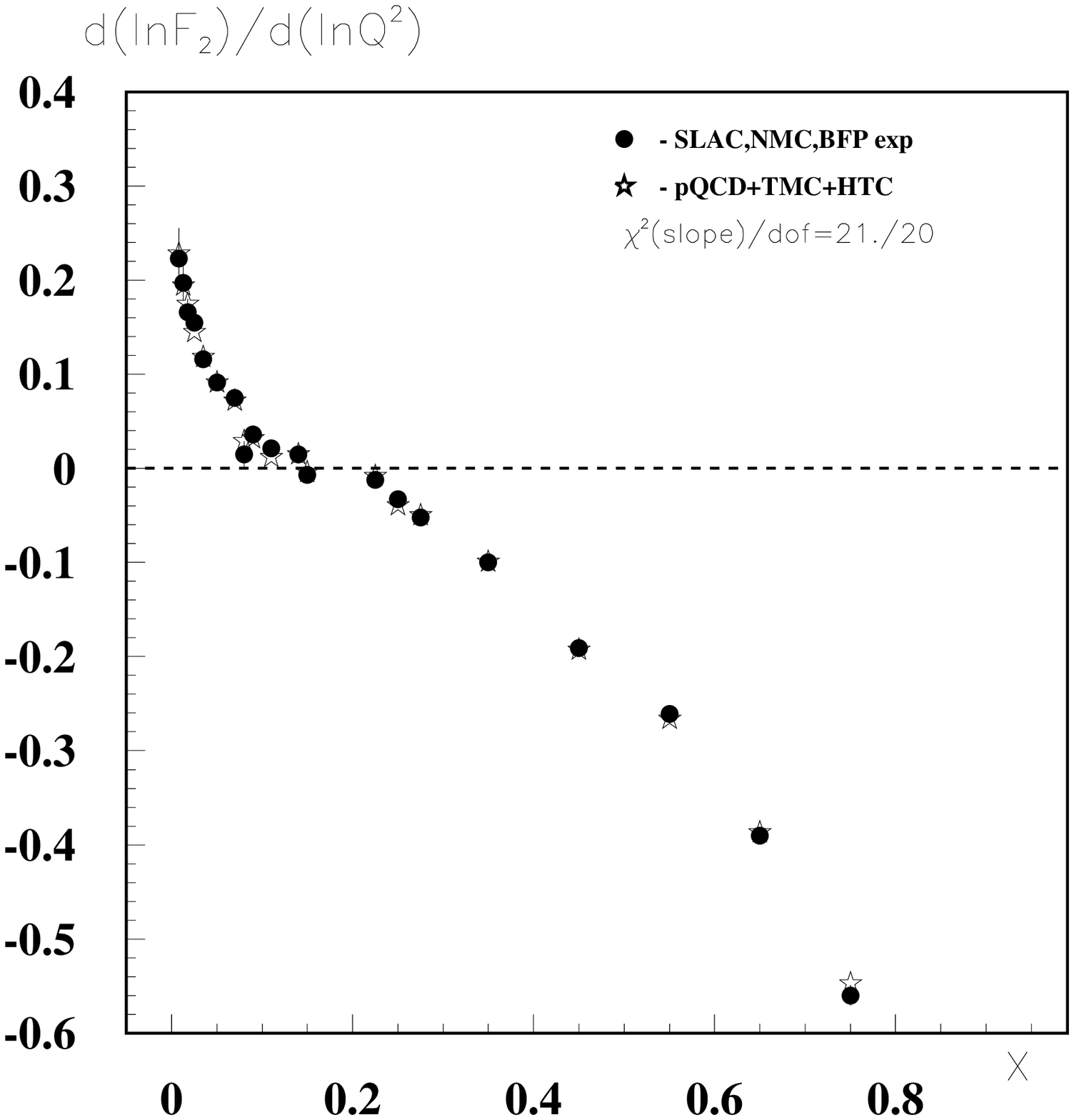,width=15cm,height=12cm}
\end{center}
\vskip -1.5cm
 \caption{
Notation as in Fig. 6 with one exception: the target mass and twist-four
corrections
%$y_{cut}$ with$N=5$ is 
are taken into account for theoretical predictions.
}
%\vskip 1cm
 \end{figure}

After the cuts have been implemented (in this Section below we use the set 
$N=5$),
we have 590 points in the analysis.
Fitting them in agreement with the same procedure considered in the 
%previous 
Section 3,
we obtain the following results:
\bea
\as(20~\mbox{GeV}^2) &=& 0.2160 \pm 0.0047 ~\mbox{(stat)} 
\pm 0.0068 ~\mbox{(syst)} 
\pm 0.0031 ~\mbox{(norm)} 
\nonumber \\
& & \label{bd1.b} \\
\as(M_Z^2) &=& 0.1175 \pm 0.0014 ~\mbox{(stat)} 
\pm 0.0020 ~\mbox{(syst)} \pm 0.0011 ~\mbox{(norm)}
\nonumber
\eea
As in the nonsinglet case the last error ($\pm 0.0011$ to $\as(M_Z^2)$) comes 
from difference in fits 
with free and fixed normalizations of BCDMS data 
\cite{BCDMS1,BCDMS2,BCDMS3} having 
different values of energy.\\

So, for the
%singlet 
fits of BCDMS data \cite{BCDMS1,BCDMS2,BCDMS3} based on complete singlet and
nonsinglet evolution  with 
minimization of
systematic errors, we have the following results (total experimental error 
is squared root of sum of squares of 
statistical error, systematic one and error of normalization):
\bea
\as(M_Z^2) ~=~ 0.1175 \pm 0.0026 ~\mbox{(total experimental error)} 
\label{bd1.b1}
\eea

The value of $\as(M_Z^2)$ corresponds to the following value 
of QCD mass parameter:
\bea
\Lambda^{(4)}_{\MSbar} &=& 
\biggl( 290 \pm 20 ~\mbox{(stat)} 
\pm 29 ~\mbox{(syst)} 
\biggr) \mbox{ MeV},\nonumber \\
\Lambda^{(5)}_{\MSbar} &=& 
\biggl( 206 \pm 17 ~\mbox{(stat)} 
\pm 24 ~\mbox{(syst)} 
\biggr) \mbox{ MeV}, 
\label{bd11.b}
\eea
where the errors connected with the type of normalization of data
are included already to systematic error.

\subsection { SLAC and NMC $H_2 + D_2$ data and BFP %iron 
$Fe$ data}

We continue our 
%singlet 
analyses with experimental data 
\cite{SLAC1,SLAC2,NMC,BFP}
obtained 
by SLAC, NM and BFP Collaborations.
The full set of data is 719 points (with the cut $Q^2>1$ GeV$^2$): 
364 ones of SLAC, 300 ones of NMC and 55 ones of BFP.
The starting point of QCD evolution is $Q^2_0=20$ GeV$^2$. 
%the $Q^2$-cut is $Q^2>1$ GeV$^2$. 

As in previous Section we give an
 illustration of importance of $1/Q^2$ corrections.
%We fit the data in the following way. 
First of all, we analyze the data 
applying only perturbative QCD part of SF $F_2$, i.e. $F_2^{tw2}$. Later, 
we add the $1/Q^2$ corrections: firstly, target mass ones and later
twist-four ones. As it is possible to see in the Table 12 and Figs. 6-8, 
we have the very bad fit ($\chi^2$(slope)$/DOF=40$), when we work only 
with twist-two part $F_2^{tw2}$.
The agreement with the data is better
%have improved 
essentially 
($\chi^2$(slope)$/DOF \approx 9$)
when target mass 
corrections have been added. The incorporation of twist-four corrections 
leads to very good fit of the 
%SLAC 
data: $\chi^2$(slope)$/DOF \approx 1.05$ (see the Table 12 and the Fig. 8) . 
We note that 
%we combine 
the statistical and systematic errors are combined in quadratures.

Thus, we see that $\chi^2$(slope)$/DOF$ decreases in 38 times when the
$1/Q^2$ corrections has been taken into account.\\

%\hskip -.56cm
%
{\bf Table 12.} The values of $\asMZ$ and $\chi^2$ at different 
regimes of fits.
\vspace{0.2cm}
\begin{center}
%\footnotesize
\begin{tabular}{|l|c|c|c|c|c|c|c|c|}
\hline
& & & & & & & \\
$N$ of  & TMC & HTC & syst. &$\chi^2(F_2)/DOF$ & $\chi^2(slope)$ 
& $\as(20~\mbox{GeV}^2)$ &
$\asMZ$ \\
fits & & & error &  &for $23$ points  &$\pm$ stat  &   \\
\hline \hline
1 & No & No &  Yes & 5.5 & 800 & 0.2400 $\pm$ 0.0017 & 
0.1241 \\  
%\hline
2 & Yes & No &  Yes & 2.2 & 179 & 0.2153 $\pm$ 0.0018 & 
0.1174 \\  
%\hline
3 & Yes & Yes & Yes & 0.85 & 21 & 0.2138 $\pm$ 0.0058 & 
0.1170 \\  
\hline
\end{tabular}
\end{center}

%\vspace{0.5cm}
\vspace{1cm}

Looking at
%carefully 
the results in the Table 12,
we see the following results for coupling constants
\bea
\as(20~\mbox{GeV}^2) &=& 0.2138 \pm 0.0058 ~\mbox{(stat)} + 
\pm 0.0075 
%\end{array} 
~\mbox{(syst)} 
+ 0.0030\mbox{(norm)}
\nonumber \\
& & \label{slo11.b} \\
%\nonumber \\
\as(M_Z^2) &=& 0.1170 \pm 0.0016 ~\mbox{(stat)} + 
%\biggl\{ \begin{array}{l}+ 0.0020 \\ - 
\pm 0.0021 
%\end{array} 
~\mbox{(syst)} 
+ 0.0011\mbox{(norm)} \nonumber
%\label{slo1}
\eea
As in the nonsinglet evolution fits,
the last error $\pm 0.0011$ to $\as(M_Z^2)$ comes from fits 
with free and fixed normalizations between  different data of SLAC,
NM and BFP Collaborations.\\

We would like to compare the results in the Table 12 with the
results of the analyses of the data when an additional $W^2$-cut
is taken into account. The inclusion of the $W^2$-cut is very popular
(see \cite{SSS} and references therein) and we fit considering data
with variation of the $W^2$-cut (and with the standard cut $Q^2>1$ GeV$^2$).
The results of the fits (without twist-four correction) are presented in 
the Table 13 (the systematic errors of the data are included in the fits).\\

%\newpage
%\hskip -.56cm
%
{\bf Table 13.} The values of $\asMZ$ and $\chi^2$ in fits with  different 
values of $W^2$-cut.
\vspace{0.2cm}
\begin{center}
%\footnotesize
\begin{tabular}{|l|c|c|c|c|c|c|c|}
\hline
& & &  & & & \\
$N$ of  & $W^2$ &$\chi^2(F_2)/DOF$ 
& $\as(20~\mbox{GeV}^2)$ & $\Lambda_{\overline{MS}}^{(4)}$ (MeV) &  
$\Lambda_{\overline{MS}}^{(5)}$ (MeV)& 
$\asMZ$ \\
fits & cut & & $\pm$ stat & $\pm$ stat & $\pm$ stat
 &$\pm$ stat     \\
\hline \hline
1 & 2.0 & 1.30 &  0.2407 $\pm$ 0.0013 & 400 $\pm$ 6 & 296 $\pm$ 4 & 
0.1243 $\pm$ 0.0004  \\  
%\hline
2 & 4.0 & 1.00 &  0.2135 $\pm$ 0.0018 & 280 $\pm$ 7 & 194 $\pm$ 5 & 
0.1169 $\pm$ 0.0004  \\  
%\hline
3 & 6.0 & 1.00 &  0.2070 $\pm$ 0.0023 & 253 $\pm$ 9 & 178 $\pm$ 7 & 
0.1150 $\pm$ 0.0007  \\  
%\hline
4 & 8.0 & 0.91 &  0.2128 $\pm$ 0.0043 & 277 $\pm$ 18 & 197 $\pm$ 14 & 
0.1167 $\pm$ 0.0012  \\  
%\hline
5 & 10 & 0.91 &  0.2107 $\pm$ 0.0053 & 268 $\pm$ 22 & 190 $\pm$ 18 & 
0.1162 $\pm$ 0.0015  \\  
%\hline
\hline
\end{tabular}
\end{center}

\vspace{0.5cm}

As we can see from the Table 12 (the last line, where twist-four corrections
are incorporated) and Table 13, the results 
for $\asMZ$ are in very good agreement for values of $W^2$-cut larger then
4 GeV$^2$. So, the $W^2$-cut procedure can be used successfully to switch off
the range of experimental data where higher-twist corrections are
%should be
required.\\

We would like to note that
%Taking together 
the results obtained in two previous subsections show
%we see 
very good agreement 
%(within existing errors) 
between the values of coupling 
constant $\as(M_Z^2)$ obtained
separately in the fits of BCDMS data and ones in the fits of combine 
SLAC, NMC and BFP data (see Eqs. (\ref{bd1.b})-(\ref{bd11.b}) and 
(\ref{slo11.b})).
Thus, as in the case of nonsinglet evolution 
we have possibility to fit togather all the data. It
is the subject of the following subsection.

\subsection { SLAC, BCDMS, NMC and BFP data }
%{\bf 4.1.} {\bf SLAC data}

Here we start to analyze the maximal number of experimental points which have
been produced in considered experiments.
The full set of data is 1309 points. \\

{\bf 5.3.1. The study of $Q^2$ range, 
where $1/Q^2$ corrections are important. }\\
%{\bf Nonsinglet case}

Here we would like to repeat our analysis given in Subsection 4.3 .
Firstly we fit the data without a contribution of twist-four terms.
We use the cut $Q^2 \geq Q^2_{cut}$ and
increase the value $Q^2_{cut}$ step by step.
Later we do  fits including the twist-four corrections and 
the cut $Q^2 \geq 1$ GeV$^2$.

As it was in nonsinglet case, we observe very good agreement for first type
of  the fits starting with  $Q^2_{cut} \geq 15$ GeV$^2$ (see the Table 14
and the Fig. 9). 
For the second
type of fits the agreement is good already at $Q^2 \geq 1$ GeV$^2$.
The both types of the fits give very similar results.
Moreover, the results are very close to ones obtained earlier in the 
nonsinglet case  (see the Tables 3 and 6).\\

%\newpage

\begin{figure}[tb]
\vskip -1cm
\begin{center}
\psfig{figure=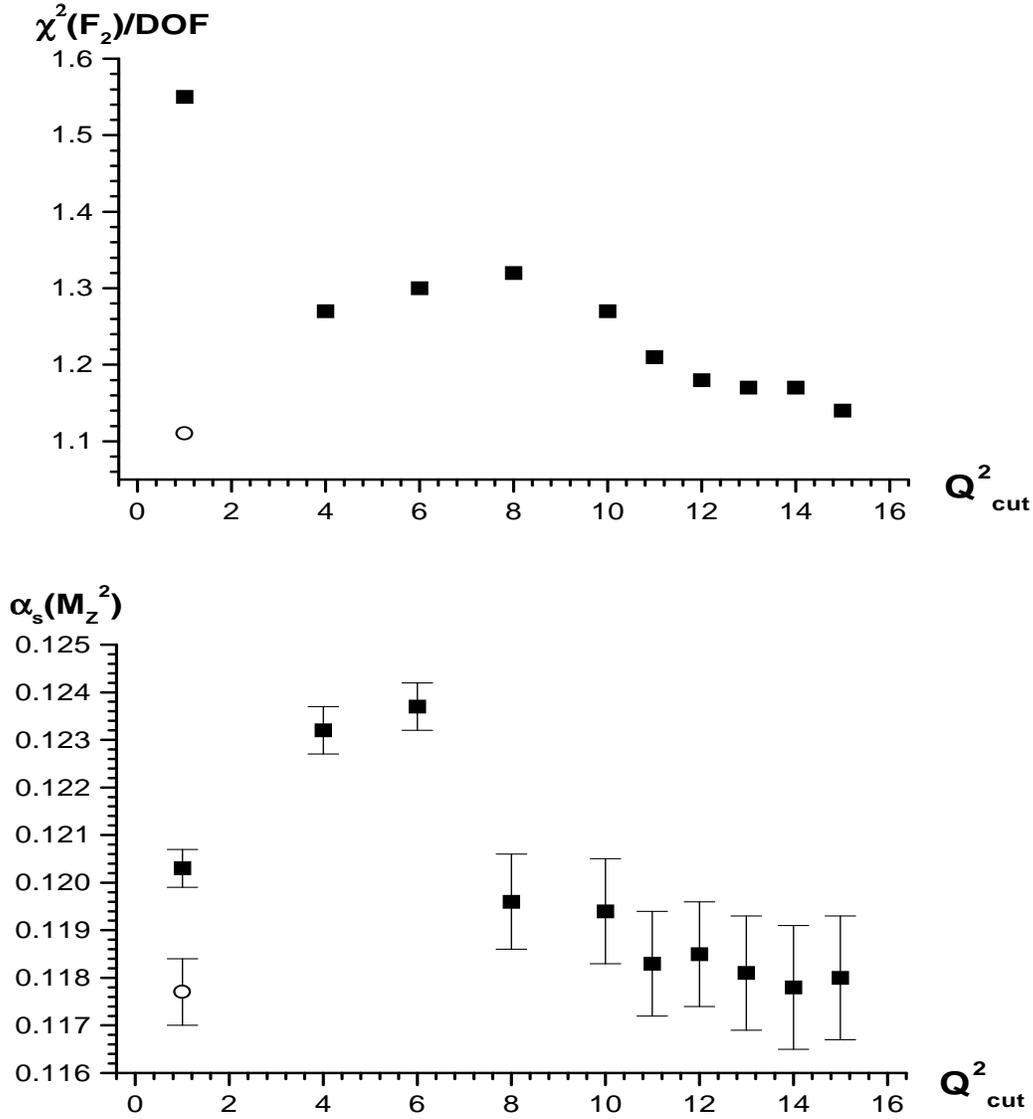,width=14cm,height=16cm}
\end{center}
\vskip -0.8cm
\caption{
The values of $\asMZ$ and $\chi^2$ at different $Q^2$-values of data cutes
in the fits based on combine singlet and nonsinglet evolution.
%regimes of fits. 
The black 
%and white 
points show the 
analyses of data without  twist-four contributions.
The white point corresponds to the case where twist-four contributions 
were added.   Only statistical errors are shown.
}
\vskip 0.3cm
\label{Non}
\end{figure}

%\newpage
%\hskip -.56cm
%
{\bf Table 14.} The values of $\asMZ$ and $\chi^2$ at different 
regimes of fits.
\vspace{0.2cm}
\begin{center}
%\footnotesize
\begin{tabular}{|l|c|c|c|c|c|c|c|}
\hline
& & & & & & & \\
$N$ of & $Q^2$ &  $N$ of & HTC &$\chi^2(F_2)$/DOF&  
$\as(20~\mbox{GeV}^2)$ & 
$\Lambda_{\overline{MS}}^{(4)}$ &  
$\asMZ$ \\
fits & cut & points & & & $\pm$ stat & (MeV) & $\pm$ stat  \\
\hline \hline
1 & 1.0 & 1309 &  No & 1.55 & 0.2258 $\pm$ 0.0011 & 333 & 
0.1203 $\pm$ 0.0004 \\  
2 & 4.0 & 1051 &  No & 1.27 & 0.2364 $\pm$ 0.0017 & 380 & 
0.1232 $\pm$ 0.0005 \\  
3 & 6.0 & 942 &  No & 1.30 & 0.2385 $\pm$ 0.0022 &  390 & 
0.1237 $\pm$ 0.0005 \\  
4 & 8.0 & 870 &  No & 1.32 & 0.2232 $\pm$ 0.0035 & 321 & 
0.1196 $\pm$ 0.0010 \\  
5 & 10.0 & 817 &  No & 1.27 & 0.2226 $\pm$ 0.0035 & 318  & 
0.1194 $\pm$ 0.0011 \\  
%\hline
6 & 11.0 & 793 &  No & 1.21 & 0.2187 $\pm$ 0.0038 &  301 &
0.1183 $\pm$ 0.0011 \\  
%\hline
7 & 12.0 & 758 &  No & 1.18 & 0.2192 $\pm$ 0.0039 & 304 &
0.1185 $\pm$ 0.0011 \\  
%\hline
8 & 13.0 & 754 &  No & 1.17 & 0.2180 $\pm$ 0.0039 & 297  &
0.1181 $\pm$ 0.0012 \\
%\hline
9 & 14.0 & 740 &  No & 1.17 & 0.2169 $\pm$ 0.0041 & 294 &
0.1178 $\pm$ 0.0013 \\ 
%\hline
10 & 15.0 & 714 &  No & 1.14 & 0.2177 $\pm$ 0.0042 & 297 &
0.1180 $\pm$ 0.0013 \\ 
\hline \hline
11 & 1.0 & 1309 &  Yes & 1.11 & 0.2167 $\pm$ 0.0024 & 293 &
0.1177 $\pm$ 0.0007 \\  
\hline
\end{tabular}
\end{center}

\vspace{0.5cm}
%\vspace{-0.2cm}

\begin{figure}[tb]
\vskip 0cm
\begin{center}
\epsfig{figure=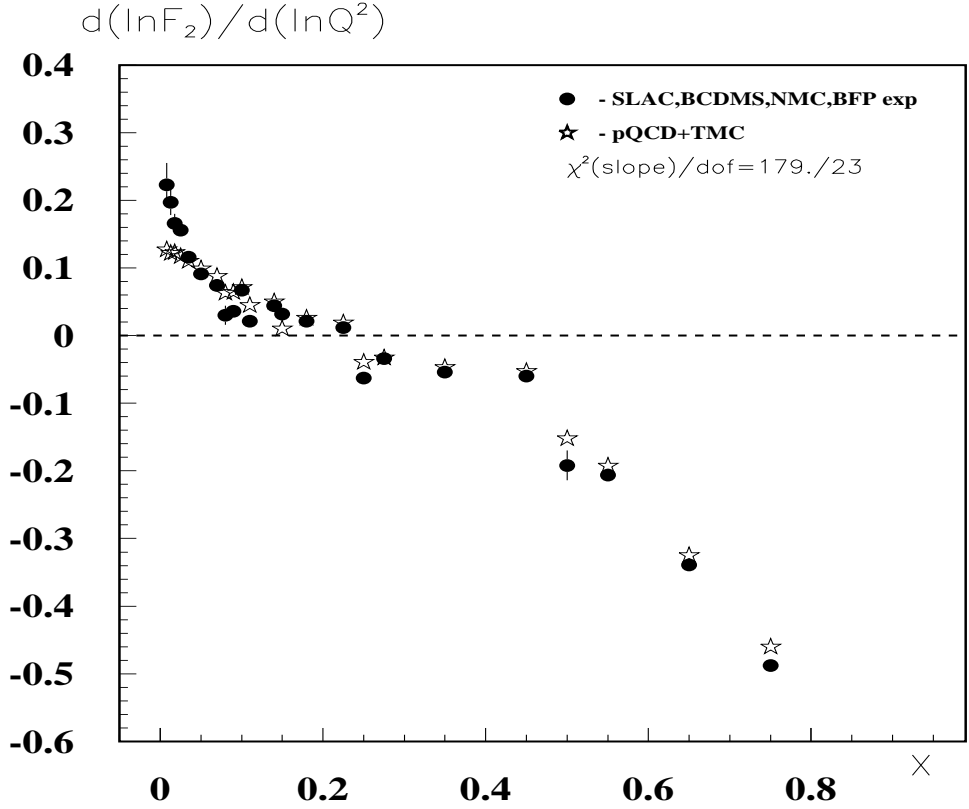,width=15cm,height=12cm}
\end{center}
\vskip -1.5cm
 \caption{
The values of the slope $d(\ln{F_2})/d(\ln{Q^2})$  at $Q^2=20$ GeV$^2$.
The white points correspond to the theoretical predictions
based on 
%the twist-two approximation of 
perturbative QCD (with target mass corrections taken into account)
and
combine singlet and nonsinglet evolution.
%at $Q^2=??$ GeV$^2$ and $Q^2_0=20$ GeV$^2$, respectively.
The black points show SLAC, BCDMS, NMC and BFP experimental data
without a $Q^2$ cut.
}
%\vskip 1cm
 \end{figure}

\begin{figure}[tb]
\vskip 0cm
\begin{center}
\epsfig{figure=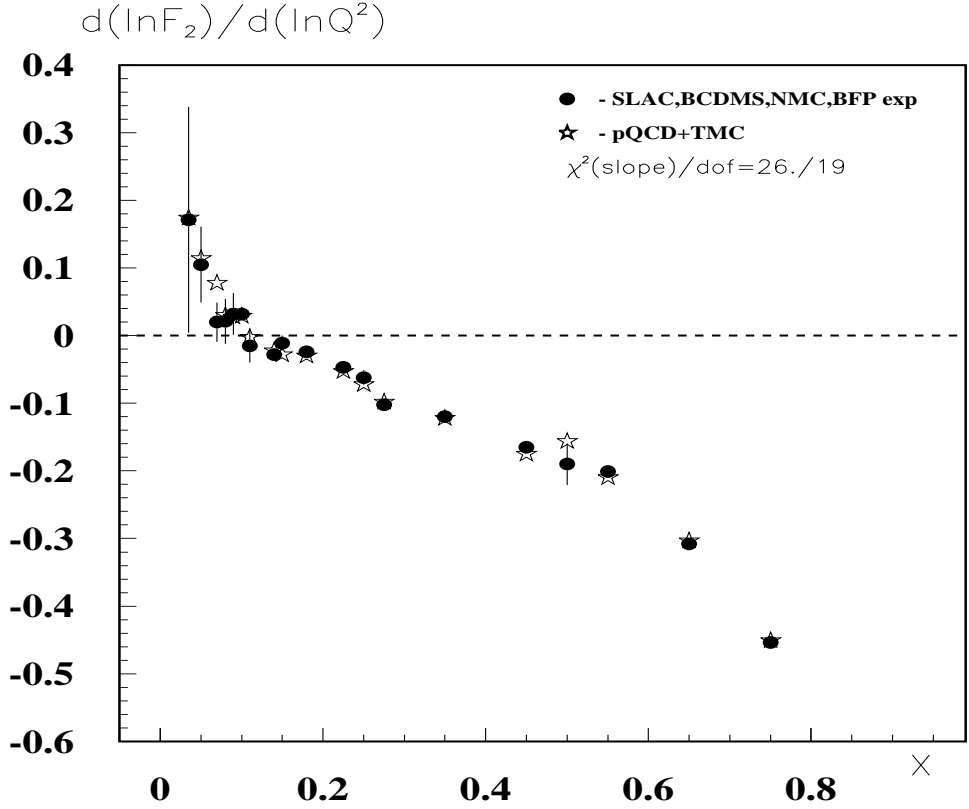,width=15cm,height=12cm}
\end{center}
\vskip -1.5cm
 \caption{
Notation as in Fig. 10 with one exception: 
the cut $Q^2 \geq 15$ GeV$^2$ is taken into account for
experimental data.
}
%\vskip 1cm
 \end{figure}

\begin{figure}[tb]
\vskip 0cm
\begin{center}
\epsfig{figure=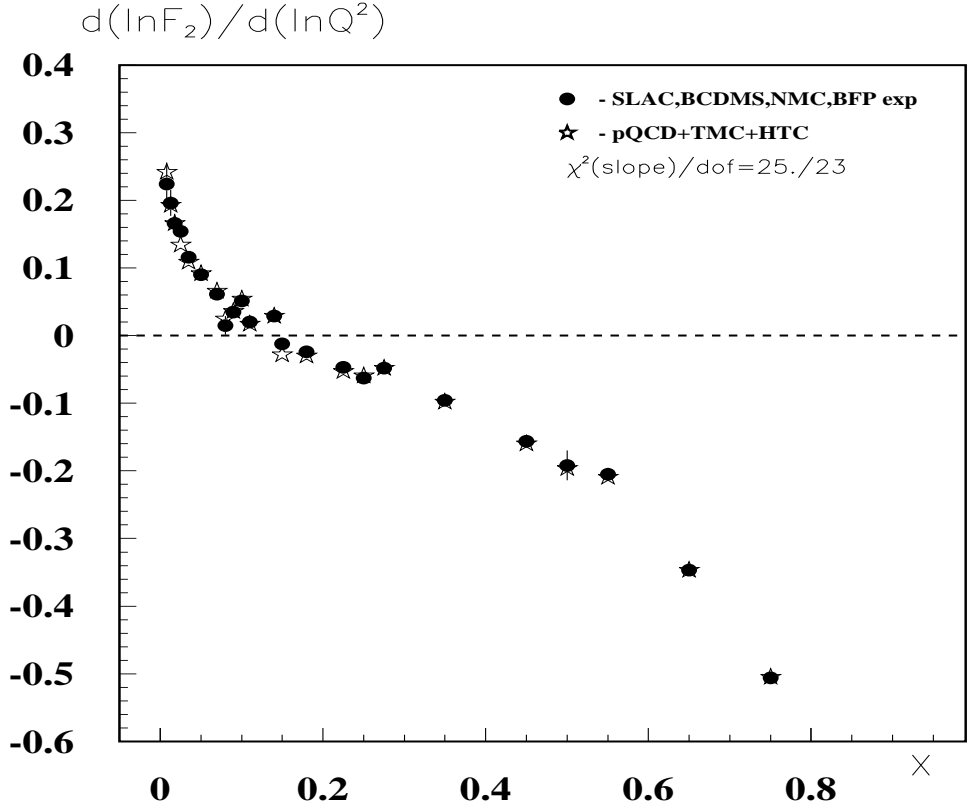,width=15cm,height=12cm}
\end{center}
\vskip -1.5cm
 \caption{
Notation as in Fig. 10 with one exception: the 
%target mass and
twist-four corrections
are taken into account for theoretical predictions.
}
%\vskip 1cm
 \end{figure}

We obtain the following results:
\begin{itemize}
\item  
When twist-four
%HT 
corrections are not included and the cut of $Q^2$ is 15 GeV$^2$ 
\bea 
\chi^2\mbox{/DOF}~=~1.14~~~\mbox{ and }~~~
\as(20~\mbox{GeV}^2) &=& 0.2177 \pm 0.0042 ~\mbox{(stat)}, \nonumber \\
\as(M_Z^2) &=& 0.1180 \pm 0.0013 ~\mbox{(stat)}
\label{fusi1}
\eea
\item  
When twist-four
%HT 
corrections are included and the cut of $Q^2$ is 1 GeV$^2$ 
\bea 
\chi^2\mbox{/DOF}~=~1.11~~~\mbox{ and }~~~
\as(20~\mbox{GeV}^2) &=& 0.2167 \pm 0.0024 ~\mbox{(stat)}, \nonumber \\
\as(M_Z^2) &=& 0.1177 \pm 0.0007 ~\mbox{(stat)}
\label{fusi2}
\eea
\end{itemize}

For additional
%another 
illustration of importance of $1/Q^2$ corrections at
nonlarge $Q^2$ values we 
%repeat of the 
study 
%of 
the slope 
$d(\ln{F_2})/d(\ln{Q^2})$ as it has been done
%analysis given 
in the previous subsection 5.2 .
First of all, we analyze the data 
applying only 
%the twist-two approximation of 
perturbative QCD approximation
%part 
of SF $F_2$ (with target mass corrections taken into account), 
i.e. $F_2^{pQCD}$. Later, 
we add the cut $Q^2 \geq 15$ GeV$^2$.
As it is possible to see in the Figs. 10 and 11, 
we have the bad fit ($\chi^2$(slope)$/DOF \approx 7.78$)
in the case without a $Q^2$ cut.
The agreement with the data 
%have improved 
is strongly better
when this $Q^2$ cut has
been added: $\chi^2$(slope)$/DOF \approx 1.26$ in the case.

As in the previous subsection
the incorporation of twist-four corrections 
leads also to very good fit of the data (without a $Q^2$ cut): 
$\chi^2$(slope)$/DOF \approx 1.09$ (see the Fig. 12) . 
These results demonstrate the importance of twist-four corrections 
at nonlarge $Q^2$ values.

Thus, as it follows from the fits of experimental data based on combine
singlet and nonsinglet evolution, 
perturbative QCD works  well at $Q^2 \geq 15$ GeV$^2$.

\vspace{0.5cm}

{\bf 5.3.2. The study of threshold effects. }

Here we continue our study of 
threshold effects in $Q^2$-evolution of SF $F_2$.
Note that at LO level and  NLO one in the singlet case of evolution 
the coefficient functions of $F_2$
and anomalous dimensions depend on the number $n_f$ of active quarks.

By analogy with the NS case of evolution (see subsection 4.3.2),
to study the threshold effects we consider two types of possible thresholds
of heavy quarks: $Q^2_f = 4m^2_f$ and $Q^2_f = m^2_f$. First type of
thresholds has appeared when a heavy quark with the mass $m_f$ takes a
possibility to be born (in the framework of photon-gluon fusion process, for
example). The second one lies close to the position
of ``Euclidean-reflected'' threshold of heavy quarks. It should play
a significant role (see \cite{SSiMi1}) in the $\as(Q^2)$-evolution.\\

{\bf A.~~ } Let thresholds appear
%take places 
at $Q^2_f=4m_f^2$. 
Then we split the range
of the data to three separate ones (see page 20).\\

%\vspace{0.5cm}

%\newpage
%\hskip -.56cm
%
{\bf Table 15.} The values of $\asMZ$ and $\chi^2$ at different 
regimes of fits.
\vspace{0.2cm}
\begin{center}
%\footnotesize
\begin{tabular}{|l|c|c|c|c|c|c|c|c|c|}
\hline
& & & & & & & & &\\
$N$ of & $Q^2$ & $n_f$ & $Q^2_0$ & $N$ of &$\chi^2$&  
$\Lambda_{\overline{MS}}^{(3)}$ & 
$\Lambda_{\overline{MS}}^{(4)}$ &  $\Lambda_{\overline{MS}}^{(5)}$ &
$\asMZ$ \\
fit & range &  &  & points &  & $\pm$ stat & $\pm$ stat & $\pm$ stat 
& $\pm$ stat \\
& &  &  &  &  & (MeV) & (MeV) & (MeV) &  \\
\hline \hline
1 & 1-10 & 3 & 3.0 & 467 & 290 & 331 $\pm$ 24 & 250 $\pm$ 20 & 176 $\pm$ 16 &
0.1148 $\pm$ 0.0015 \\  
%\hline
2 & 10-80 & 4 & 20 & 627 & 595 & & 274 $\pm$ 21 & 194 $\pm$ 17 &
0.1165 $\pm$ 0.0014 \\  
%\hline
3 & 80-300 & 5 & 90 &190 & 156 & & & 220 $\pm$ 70 &
0.1187 $\pm$ 0.0050 \\  
\hline
\end{tabular}
\end{center}

\vspace{0.5cm}
%\vspace{1cm}

The results are shown in Table 15.
The average $\asMZ$ value can be calculated
and it has the following
value:
\bea 
\as(M_Z^2) &=& 0.1158 \pm 0.0010 ~\mbox{(stat)}
\label{thres.1}
\eea

\vspace{0.5cm}

{\bf B.~~ } Let thresholds appear
%take places 
at $Q^2_f=m_f^2$. 
Then we split the range of the data to two separate ones (see page 20).

\vspace{0.5cm}

{\bf Table 16.} The values of $\asMZ$ and $\chi^2$ at different 
regimes of fits.
\vspace{0.2cm}
\begin{center}
\begin{tabular}{|l|c|c|c|c|c|c|c|c|}
\hline
& & & & & & &  &\\
$N$ of & $Q^2$ & $n_f$ & $Q^2_0$ & $N$ of &$\chi^2$&   
$\Lambda_{\overline{MS}}^{(4)}$ & $\Lambda_{\overline{MS}}^{(5)}$ &
$\asMZ$ \\
fit & range &  &  & points &  & $\pm$ stat & $\pm$ stat & $\pm$ stat  \\
& &  &  &  &  & (MeV) & (MeV) &   \\
\hline \hline
1 & 2.5-20.5 & 4 & 10 & 519 & 396 & 230 $\pm$ 21 & 160 $\pm$ 16 &
0.1132 $\pm$ 0.0016 \\  
2 & 20.5-300 & 5 & 90 & 631 & 670 & & 205 $\pm$ 15 &
0.1174 $\pm$ 0.0013 \\  
\hline
\end{tabular}
\end{center}

\vspace{0.5cm}

The results are shown in Table 16.
The average $\asMZ$ value can be calculated 
and it has the following
value:
\bea 
\as(M_Z^2) &=& 0.1157 \pm 0.0020 ~\mbox{(stat)}
\label{thres.2}
\eea

The results are very surprising.
%interested.
From one side, all variables: the coefficient functions of $F_2$
and anomalous dimensions, depend on the number $n_f$ of active quarks.
However, we do not find a strong dependence on exact value of thresholds
of heavy quarks. 
From another side, the central values of the average $\asMZ$  obtained 
here are essentially lower than in other our analyses.\\

Thus, the theoretical uncertainties due to threshold effects can be 
estimated in the case of combine singlet and nonsinglet evolution for
$\as(M_Z^2)$ as $0.0001$.\\

{\bf 5.3.3. The values of fitted parameters. }\\

%\end{document}
\begin{figure}[tb]
\vskip -0.5cm
\begin{center}
\psfig{figure=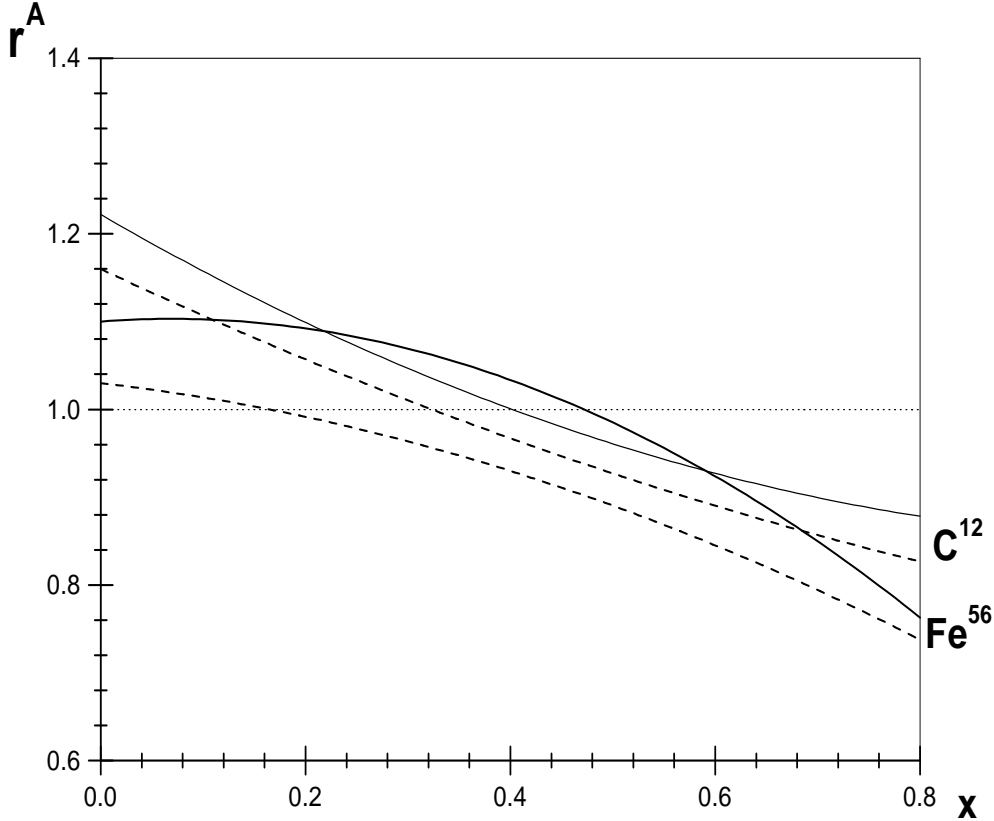,width=16cm,height=13cm}
\end{center}
\vskip -1cm
\caption{
The values of the nuclear-effect ratio: $r^A$, for $A=C^{12}$ and $Fe^{56}$. 
The solid and dashen curvess correspond 
to the small-$x$ asymptotics $\sim x^{-\omega}$ of sea quark and gluon
distributions with $\omega =0$ and
$\omega =0.18$, respectively. 
%The statistical errors are displayed. 
}
\label{nucl}
\end{figure}

We have got the following values for parameters in parameterizations of
parton distributions (at $Q^2_0=20$ GeV$^2$)
\footnote{Here and in the following subsection we give the results for
the coefficient $P_G(Q^2_0)$ but not for the one $C_G(Q^2_0)$. They are
connected because of Eq. (\ref{4.3a}):
$P_G(Q^2_0)=C_G(Q^2_0)\cdot B(a_g(Q^2_0)+1,b_g(Q^2_0)+1)$, where 
the beta-function $B(a,b)$ has been defined in Eq.(\ref{4.2v}).}:
\bea 
a_u(20) &=& 0.72, 
%\pm 0.02
~~~~~~ b_u(20)~=~3.72, 
%\pm 0.02,  
\nonumber \\
a_d(20) &=& 0.69, 
%\pm 0.02,
~~~~~~ b_d(20)~=~5.81,
% \pm 0.02,  
\nonumber \\
C_S(20) &=& 0.375,
% \pm 0.009,
~~~~\, b_S(20)~=~13.8, 
%\pm 0.4,  
\nonumber \\
P_G(20) &=& 0.519, 
%\pm 0.002,
~~~~\, b_G(20)~=~11.4, 
%\pm 0.9,  
\nonumber \\
K_1^C(20) &=& 1.222,
% \pm 0.008,
~~~ K_2^C(20) ~=~ 0.554, 
%\pm 0.019,
~~~~\, K_3^C(20) ~=~ 0.253,
% \pm 0.020,
  \nonumber \\
K_1^F(20) &=& 1.10,
% \pm 0.01,
~~~~\, K_2^F(20) ~= -0.081,
% \pm 0.123,
~~\, K_3^F(20) ~= -0.58 
%\pm 0.21
\label{para2}
\eea

For the coefficients $a_u(20)$ and $a_d(20)$
%$b_{NS}^l$ ($l=P,D,F$) 
we find good agreement between their values and the
double-logarithmic estimations in Refs.\cite{Ermolaev,Ermolaev1},
based on \cite{KiLi}. We would like to note that the estimations
in Ref. \cite{Ermolaev} have been given in other set of parameters that 
changes effectively only the value of normalization point $Q^2_0$.
As it has been shown in 
%quite old articles 
Refs. \cite{LoYn,VoKoMa},
the value of $a_u(20)$ and $a_d(20)$
%$b_{NS}^l$ 
should be nearly $Q^2$-independent (if the
%$b_{NS}^l$ 
values are not too close to $1$) 
%or $-1$)
\footnote{This $Q^2$-independence is very similar to corresponding 
$Q^2$-independence of the coefficients $a_S(20)$ and $a_G(20)$
in the power-like
small $x$ asymptotics $\sim x^{a_S}$ and $\sim x^{a_G}$ of singlet 
parton distributions, if $a_S$ and $a_G$ are not close numerically to $0$ 
%or $-1$
(see studies in Refs.\cite{LoYn1,KoMaPa,Ko94,Ko96,KoPaGFL} and references
therein).}.
%, as it follows from BFKL analysis \cite{BFKL}.}.
This $Q^2$-independence of 
%$b_{NS}^l$ 
values of $a_u(20)$ and $a_d(20)$  
explains our good agreement with the results of \cite{Ermolaev}.
%As we have already mentioned, 
The 
values of $a_u(20)$ and $a_d(20)$ are supported also 
by recent fits (see discussions in 
Ref. \cite{KPS1}). 

The value $b_u(20)$ is 
%also 
in agreement
with Eqs.(\ref{para}) and (\ref{para1}) and with other fits 
\cite{Al2000,Al2001,KPS,KPS1}, that supports its slow $Q^2$-dependence
(see \cite{Gross,VoKoMa}). The value of $b_d(20)$ is higher than
$b_u(20)$ that is supported by
%is in agreement with 
other fits (see, for example, 
\cite{Al2000,Al2001}) and references therein) and by quark-counting rules
\cite{schot}. The values of $b_G(20)$ and $b_S(20)$ are very high, that is
in agreement with BCDMS analyses \cite{BCDMS1,BCDMS2,BCDMS3} and demonstrates
difficulties to study the large-$x$ asymptotics of sea quark and gluon
%parton 
distributions
in analyses of inclusive deep-inelastic data
\footnote{ In semi-inclusive case of deep-inelastic scattering the gluons
give  large contributions, essentially at low $x$ values, 
(see, for example, the 
recent study of open charm production
in \cite{KoLiPaZo} and references therein) and, thus, gluon
distribution can be perfectly extracted.}.
 
\begin{figure}[tb]
\vskip -0.5cm
\begin{center}
\psfig{figure=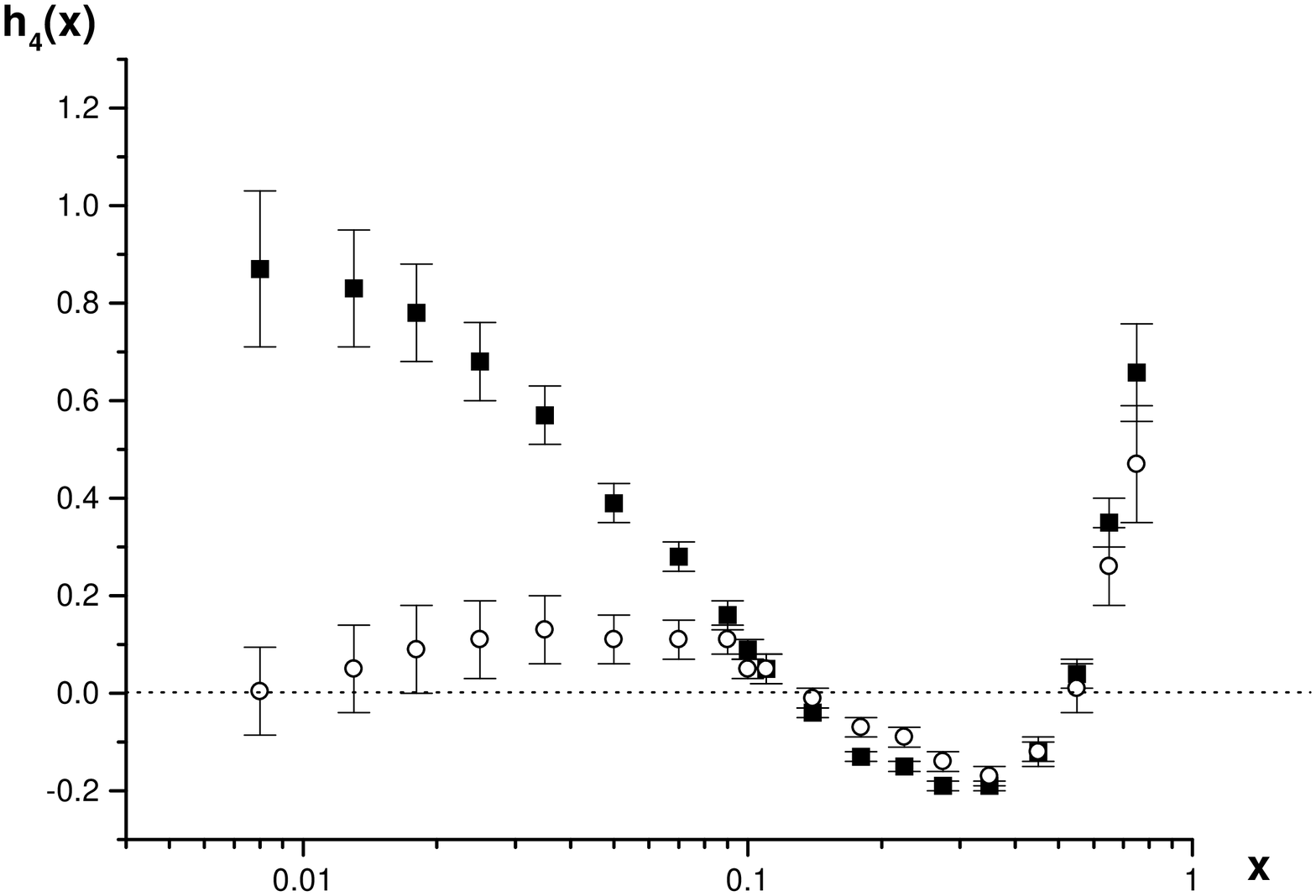,width=16cm,height=13cm}
\end{center}
\vskip -1cm
\caption{
The values of the twist-four terms. The black and white points correspond 
to the small-$x$ asymptotics $\sim x^{-\omega}$ of sea quark and gluon
distributions
with $\omega =0$ and
$\omega =0.18$, respectively. The statistical errors are displayed only. 
}
\label{HT}
\end{figure}

The value of $P_G(20)$ shows that at $Q^2_0=20$ GeV$^2$ gluons contain 
about half of nucleon momentum.

The coefficients $K^C_i(20)$ and $K^F_i(20)$ ($i=1,2,3$) demonstrate non-zero 
values of nuclear effects for bound nucleons in $C^{12}$ and $Fe$ nuclei.
The Eqs. (\ref{nu3}) with the values of coefficients $K^C_i(20)$ and 
$K^F_i(20)$
given in Eqs. (\ref{para2}) demonstrate the shapes of nuclear effects which
are represented in Fig 3, where we see a resonable
%good 
agreement of our curves with the 
experimental data from Refs \cite{EMCr,BCDMSr}.\\

The values of twist-four terms are given in the Table 17. To obtain
the values we used the approximate equality of twist-four terms
for $H_2$ and $D_2$ targets that has been obtained in our studies
in the previous Section (see the Tables 5 and 7). This is also
in agreement with the Ref. \cite{ViMi}. The values of twist-four
terms are represented also in Fig. 14.

We would like to note (see the Table 17 and Fig. 14)
about a quite strong rise of twist-four terms at 
lower $x$-bins. 
The necessity of large magnitude of twist-four corrections at the low
$x$ values it is possible to observe also in the Figs. 6, 7 and 10,
where there is a quite strong difference between experimental data
and theoretical predictions (based on perturbative QCD) for the slope
$d(\ln{F_2})/d(\ln{Q^2})$.
%We note that the 
The rise is in good agreement
with theoretical predictions \cite{Bartels} and 
%in agreement 
with the
recent analyses of H1 and ZEUS data at low values of $x$ and $Q^2$
(see \cite{HT}). \\

%\vspace{0.5cm}

{\bf Table 17.} The values of the twist-four terms.
%$\asMZ$ and $\chi^2$ at different regimes of fits.
\vspace{0.2cm}
\begin{center}
%\footnotesize
\begin{tabular}{|l|c||l|c||l|c|c|}
\hline
& & & &  & \\
$x_i$  & $\tilde h_4(x_i)$ &  
$x_i$  & $\tilde h_4(x_i)$ &   
$x_i$  & $\tilde h_4(x_i)$ \\
 &   $\pm$ stat & & $\pm$ stat &  & $\pm$ stat    \\
\hline \hline
0.008 & 0.87 $\pm$ 0.16  &  0.090 & 0.16 $\pm$ 0.03  & 0.275 & -0.19 
$\pm$ 0.01 \\  
%\hline
0.013 & 0.83 $\pm$ 0.12 &  0.100 & 0.09 $\pm$ 0.02  & 0.350 & -0.19 
$\pm$ 0.01 \\  
%\hline
0.018 & 0.78 $\pm$ 0.10 &  0.110 & 0.05 $\pm$ 0.03  & 0.450 & -0.12 
$\pm$ 0.02\\  
%\hline
0.025 & 0.68 $\pm$ 0.08 &  0.140 & -0.04 $\pm$ 0.01 & 0.500 & 0.45 $\pm$
0.23 \\  
%\hline
0.035 & 0.57 $\pm$ 0.06 &  0.150 & 0.43 $\pm$ 0.11 &  0.550 & 0.04 $\pm$ 
0.03 \\  
%\hline
0.050 & 0.39 $\pm$ 0.04 &  0.180 & -0.13 $\pm$ 0.01 & 0.650 & 0.35 $\pm$ 
0.05 \\  
%\hline
0.070 & 0.28 $\pm$ 0.03 &  0.225 & -0.15 $\pm$ 0.01 & 0.750 & 0.66 $\pm$ 
0.10 \\  
%\hline
0.080 & 0.30 $\pm$ 0.15 &  0.250 & -0.27 $\pm$ 0.13 &   &  \\  
\hline
\end{tabular}
\end{center}

%\vspace{0.5cm}
\vspace{1.0cm}

{\bf 5.3.4. BFKL-like parameterizations of 
%singlet parton 
gluon and sea quark distributions. }\\

As we have already discussed in Section 2, we would like to try to study
the parameters of the sea quark and gluon distributions when the terms
$\sim x^{a_S(Q^2_0)}$ and $\sim x^{a_G(Q^2_0)}$ were incorporated. 
These terms
%which can be 
take into
account a possible rise of the sea quark and gluon distributions
at low $x$ values. As it has been already noted
in Section 2, from DGLAP-like analyses \cite{KoMaPa,Ko96,JeKoPa}, the
parameters $a_S$ and $a_G$ should be the same, because they are mixed together
into the ``$+$''-component of the $Q^2$-evolution (see \cite{Ko96}). Moreover,
the parameter $\omega= - a_S=- a_G$ should be $Q^2$-independent 
(see, for example, \cite{KoMaPa,JeKoPa}), if it is
not small, i.e. $x^{-\omega} >> Const$ at small $x$. \\

In the fit with free nonzero $\omega$ value 
we have got the following values for parameters in parameterizations of
parton distributions (at $Q^2_0=20$ GeV$^2$)
\footnote{ We would like to note that the fit contains strong correlations
between the values of $\omega$, the coupling constant and twist-four
terms. These correlations come because of very limited numbers of 
experimental data used here lie at the low $x$ region. Indeed, only the NMC
experimental data contribute there. Then, the results (\ref{para3}) can be
considered seriously only when H1 and ZEUS data \cite{H1,ZEUS} have been
taken into account. We hope to incorporate the HERA data \cite{H1,ZEUS}
in our future investigations.}:
\bea 
a_u(20) &=& 0.72,
% \pm 0.02 ,
~~~~~~~ b_u(20)~=~3.69,
% \pm 0.02,  
\nonumber \\
a_d(20) &=& 0.68,
% \pm 0.02,
~~~~~~~ b_d(20)~=~5.44,
% \pm 0.02,  
\nonumber \\
a_S(20) &=& -0.18,
% \pm 0.02 ,
~~~~ a_G(20)~=~-0.18,
% \pm 0.02,  
\nonumber \\
C_S(20) &=& 0.185,
% \pm 0.009,
~~~~~\, b_S(20)~=~10.4,
% \pm 0.4,  
\nonumber \\
P_G(20) &=& 0.524,
% \pm 0.002,
~~~~~ b_G(20)~=~7.31,
% \pm 0.86, 
 \nonumber \\
K_1^C(20) &=& 1.160,
% \pm 0.008,
~~~\, K_2^C(20) ~=~ 0.472,
% \pm 0.019,
~~\,K_3^C(20) ~=~ 0.141,
% \pm 0.020,
  \nonumber \\
K_1^F(20) &=& 1.03,
% \pm 0.01,
~~~~~ K_2^F(20) ~=~ 0.131,
% \pm 0.123,
~~\, K_3^F(20) ~=~ -0.28 
%\pm 0.21
\label{para3}
\eea

We would like to note that the values of parameters of valent quark 
distributions are not changed really. The values of $b_G(20)$ and
$b_S(20)$ are yet high but they are closer to predictions of
quark-counting rules \cite{schot} than the corresponding 
values obtained in the previous subsection.

The values of the parameters of the nuclei effect ratio  are not changed 
%also 
within considered errors. The similarity of the results for the
nuclei effect ratio is shown in the Fig. 13.

The value of $\omega$ is equal to $0.18$, that is in 
perfect agreement with the
recent studies based on BFKL dynamics \cite{BFKL} when NLO corrections
%at LO$\&$NLO approximation 
\cite{BFKLnlo,KoLi} were taken into account
(see, for example, studies \cite{BFKLP}, a review \cite{BFKLrew} 
and references therein). Moreover,
this value is in good agreement also with recent phenomenological studies
(see a recent review in \cite{Kaidalov}) of Pomeron intercept values
and also with recent H1 and L3 data \cite{H1,L3}. 

As it is possible to see in the Tables 17 and 18 and also in Fig. 14,
the effect of strong rise
of twist-four magnitude at small $x$ values observed in previous subsection
is completely absent here
\footnote{ As it was  in previous subsection, to obtain
the values we used the approximate equality of twist-four terms
for $H_2$ and $D_2$ targets that have been obtained in our studies
in the previous Section (see the Tables 5 and 7). This is also
in agreement with the Ref. \cite{ViMi}.}. 
So, the rise is replaced by the small $x$ rise of twist-two
%singlet parton 
gluon and sea quark distributions. This replacement seems due to
a small number of experimental points at low $x$ range and narrow range 
of $Q^2$ values there. The cancellation
of twist-four corrections at low $x$ is in good agreement with the recent 
studies \cite{Q2evo,Bartels1}. This demonstrates the fact that a strong
rise of twist-four corrections coming from BFKL-like approaches \cite{Bartels}
has negligible magnitude (see \cite{Bartels1,HT}).\\

%\vspace{0.5cm}

{\bf Table 18.} The values of the twist-four terms.
%$\asMZ$ and $\chi^2$ at different regimes of fits.
\vspace{0.2cm}
\begin{center}
%\footnotesize
\begin{tabular}{|l|c||l|c||l|c|c|}
\hline
& & & &  & \\
$x_i$  & $\tilde h_4(x_i)$ &  
$x_i$  & $\tilde h_4(x_i)$ &   
$x_i$  & $\tilde h_4(x_i)$ \\
 &  & $\pm$ stat & $\pm$ stat &  & $\pm$ stat    \\
\hline \hline
0.008 & 0.004 $\pm$ 0.090  &  0.090 & 0.11 $\pm$ 0.03  & 0.275 & -0.14 
$\pm$ 0.02 \\  
%\hline
0.013 & 0.05 $\pm$ 0.09 &  0.100 & 0.05 $\pm$ 0.02  & 0.350 & -0.17
$\pm$ 0.02 \\  
%\hline
0.018 & 0.09 $\pm$ 0.09 &  0.110 & 0.05 $\pm$ 0.03  & 0.450 & -0.12 
$\pm$ 0.03\\  
%\hline
0.025 & 0.11 $\pm$ 0.08 &  0.140 & -0.01 $\pm$ 0.02 & 0.500 & 0.43 $\pm$
0.23 \\  
%\hline
0.035 & 0.13 $\pm$ 0.07 &  0.150 & 0.62 $\pm$ 0.12 &  0.550 & 0.01 $\pm$ 
0.05 \\  
%\hline
0.050 & 0.11 $\pm$ 0.05 &  0.180 & -0.07 $\pm$ 0.02 & 0.650 & 0.26 $\pm$ 
0.08 \\  
%\hline
0.070 & 0.11 $\pm$ 0.04 &  0.225 & -0.09 $\pm$ 0.02 & 0.750 & 0.47 $\pm$ 
0.12 \\  
%\hline
0.080 & 0.31 $\pm$ 0.16 &  0.250 & -0.16 $\pm$ 0.14 &   &  \\  
\hline
\end{tabular}
\end{center}

\vspace{0.5cm}
%\vspace{1.0cm}

The value of  $\as(M_Z^2)$ in the fit (with the number of points $N=1309$
and $\chi^2/DOF =1.1$) is as follows: 
\bea
\as(M_Z^2) ~=~ 0.1187 \pm 0.0015 ~\mbox{(stat)}, 
%\pm 0.00?? ~\mbox{(syst)}  \pm 0.000? ~\mbox{(norm)}, 
\label{teo4a}
%\nonumber
\eea
i.e. 
%in the range of errors 
it is in good agreement within statistical errors
with fits performed earlier
but the middle value is slightly higher.

\subsection {The results of analyses with combine singlet and nonsinglet
evolution  }

Thus, using singlet analyses of the SLAC, NMC, BCDMS and BFP experimental data 
for SF $F_2$ we obtain for $\asMZ$ the following expression:

\bea
\as(20~\mbox{GeV}^2) &=& 0.2167 \pm 0.0024 ~\mbox{(stat)} 
\pm 0.0080 ~\mbox{(syst)}  \pm 0.0012 ~\mbox{(norm)} 
\nonumber \\
& & \label{teo3} \\
\as(M_Z^2) &=& 0.1177 \pm 0.0007 ~\mbox{(stat)} 
\pm 0.0021 ~\mbox{(syst)}  \pm 0.0005 ~\mbox{(norm)} 
%\label{teo4}
\nonumber
\eea

\vspace{0.5cm}

Looking at the results obtained in the Sections
we see very good agreement 
%(within existing errors) 
between the value of coupling 
constant $\as(M_Z^2)$ obtained
in the fits of combine SLAC, BCDMS, NMC and BFP data 
and the values of $\as(M_Z^2)$
obtained separately in the fits of 
BCDMS data and in ones of SLAC, BCDMS, NMC and BFP data.

\section{The dependence 
%of coupling constant 
on factorization and renormalization scales }

In the section we study the dependence of our results on the different
choice of the factorization
scale $\mu_F$ and the renormalization one $\mu_R$. 
Following  the studies \cite{ViMi,scheme}
we choose three
following values ($1/2,~1,~2$) for the coefficients $k_F$ and $k_R$.

\subsection{ Nonsinglet evolution case }

The results are given in the Table 19. We do fits here without higher-twist 
corrections (no HTC), with the number of points  $596$, at
$Q^2 > 10.5$ GeV$^2$ and for free normalization of different sets of data.
The change of the value of coupling constant $\asMZ$ at some $k_F$
and $k_R$ values is denoted by the difference:
\bea
\Delta \as(M_Z^2) ~=~ \as(M_Z^2) - \as(M_Z^2)|_{k_F=k_R=1}
\label{diff}
\eea

\vspace{0.5cm}

{\bf Table 19.} The values of $\asMZ$ at different values of $k_F$ and $k_R$.
The values in brackets correspond to the case when the Eq.(\ref{2.cf0})
replaces the Eq.(\ref{2.cc0}) into the NLO corrections to coefficient 
functions.  
\vspace{0.2cm}
\begin{center}
%\footnotesize
\begin{tabular}{|c|c||c|c|c|c|}
%\begin{tabular}{|p{35pt}|p{38pt}|p{38pt}|p{38pt}|p{38pt}|p{38pt}|p{38pt}|}
\hline
& & & & & \\
$k_R$ & $k_F$.  & $\chi^2(F_2)$ & $\as(90~\mbox{GeV}^2)$  $\pm$ stat & 
$\asMZ$  
%$\pm$ stat.err. 
& $\Delta \as(M_Z^2)$ \\
& & & & & \\
\hline \hline
1 & 1 & 556 & 0.1789  $\pm$ 0.0023 &  0.1175 
%$\pm$ 0.0012 
& 0 \\  
%\hline
1/2 & 1 & 558 & 0.1769 $\pm$ 0.0022 & 0.1167 
%$\pm$ 0.0020 
& $-$0.0008 \\  
 &  &  & (0.1745) & (0.1155) & ($-$0.0020) \\  
%\hline
1 & 1/2 & 545 & 0.1730  $\pm$ 0.0021 & 0.1150 
%$\pm$ 0.0025 
& $-$0.0025 \\  
% & &  & (0.1730)   & (0.1150)  & ($-$0.0025) \\  
%\hline
1 & 2 & 568 & 0.1876 $\pm$ 0.0025 & 0.1211 
& +0.0036 \\  
%& & & (0.1883) & (0.1210) & (+0.0035) \\  
%\hline
2 & 1 & 555 & 0.1826 $\pm$ 0.0025 & 0.1191 
%$\pm$ 0.0028 
& +0.0016 \\  
 & &  & (0.1858) & (0.1203)  & (+0.0028) \\  
%\hline
1/2 & 2 & 570 & 0.1856 $\pm$ 0.0026 & 0.1203 
%$\pm$ 0.0011 
& +0.0028 \\  
& & & (0.1817) & (0.1186) & (+0.0011) \\  
%\hline
2 & 1/2 & 554 & 0.1770 $\pm$ 0.0022 & 0.1167 
%$\pm$ 0.0017 
& $-$0.0008 \\  
& &  & (0.1784) & (0.1173) & ($-$0.0002) \\  
%\hline
1/2 & 1/2 & 556 & 0.1789 $\pm$ 0.0023 & 0.1175 
%$\pm$ 0.0011 
& $-$0.0034 \\  
& &  & (0.1694) & (0.1134) & ($-$0.0041) \\  
%\hline
2 & 2 & 567 & 0.1912 $\pm$ 0.0028 & 0.1225 
%$\pm$ 0.0023 
& +0.0050 \\ 
& & & (0.1965) & (0.1245) & (+0.0070) \\  
\hline
\end{tabular}
\end{center}

\vspace{0.5cm}

We find similar variation of $\asMZ$ with the variations of $k_F$ and $k_R$:
$\asMZ$ increases (falls) with increasing (decreasing) of values of
$k_F$ and/or $k_R$. So, the dependence
is quite similar to one which has been obtained in \cite{NeVo,KPS,KPS1} by 
the variation of $k$-scale from $1/4$ to $4$ ($k\equiv k_F=k_R$ in 
\cite{NeVo,KPS,KPS1}).

Taking maximal and minimal values (that corresponds to 
$k_R=k_F=1/2$ and $2$, respectively) of coupling constant 
we obtain the
%following 
theoretical uncertainties $+0.0050$ and $-0.0034$ for $\asMZ$. In the
case when the replacement (\ref{2.cf0}) 
has been used also in NLO corrections
to the coefficient functions (i.e. when the Eq.(\ref{2.cf0})
replaces the Eq.(\ref{2.cc0}) there), 
the theoretical uncertainties for $\asMZ$ are little higher:
$+0.0070$ and $-0.0041$.\\ 

Thus, using the analyses with NS evolution
of the SLAC, NMC, BCDMS and BFP experimental data 
for SF $F_2$ we obtain for 
$\asMZ$ the following expressions (when no HTC, $Q^2 > 10$ GeV$^2$ and  
$\chi^2=0.98$):
\bea
\as(M_Z^2) ~=~ 0.1170 \pm 0.0009 ~\mbox{(stat)} 
\pm 0.0019 ~\mbox{(syst)}  \pm 0.0010 ~\mbox{(norm)} 
+~ \biggl\{ 
\begin{array}{l}+ 0.0050 \\ - 0.0034 \end{array} ~\mbox{(theor)},
\label{teo}
\eea
or
\bea
\as(M_Z^2) ~=~ 0.1170 \pm 0.0023 ~\mbox{(total experimental error)} 
 +~ \biggl\{ 
\begin{array}{l}+ 0.0050 \\ - 0.0034 \end{array} ~\mbox{(theor)},
\label{teo1}
\eea
where
the symbol {\it theor} marks the theoretical uncertainties which
contain the sum of the scale uncertainties, threshold error ($\pm 0.0002$) 
and the method error ($\pm 0.0002$) in quadratures.

\subsection{ Combine singlet and nonsinglet evolution }

The results are given in the Table 20. We do fits with higher-twist 
corrections, with the number of points  $1309$, at
$Q^2 > 1$ GeV$^2$ and for free normalization of different sets of data.\\

{\bf Table 20.~} The values of $\asMZ$ at different values of $k_F$ and $k_R$.
The values in brackets correspond to the case when the Eq.(\ref{2.cf0})
replaces the Eq.(\ref{2.cc0}) into the NLO corrections to coefficient 
functions.
\vspace{0.2cm}
\begin{center}
%\footnotesize
\begin{tabular}{|c|c||c|c|c|c|c|c|}
%\begin{tabular}{|p{35pt}|p{38pt}|p{38pt}|p{38pt}|p{38pt}|p{38pt}|p{38pt}|}
\hline
& & & & & & & \\
$k_R$ & $k_F$.  & $\chi^2(F_2)$ & $\as(20~\mbox{GeV}^2)$  $\pm$ stat & 
$\Lambda^{(4)}_{\MSbar}$ & $\Lambda^{(5)}_{\MSbar}$ & 
$\asMZ$ & $\Delta \asMZ $   \\
& & & & MeV & MeV  & & \\
\hline \hline
1 & 1 & 1410 & 0.2167  $\pm$ 0.0024 & 293 & 209 & 0.1178 & 0 \\  
1/2 & 1 & 1410 & 0.2112 $\pm$ 0.0019 & 270 & 191 & 0.1162 & $-$0.0016 \\  
 &  & (1443)& (0.2104 $\pm$ 0.0029) & (267) & (189) & (0.1160) & 
($-$0.0018) \\  
%\hline
1 & 1/2 & 1423 & 0.2040  $\pm$ 0.0020 & 241 & 168 & 0.1140 & $-$0.0038 \\  
%\hline
1 & 2 &  1447 & 0.2300 $\pm$ 0.0031 & 351 & 256 &  0.1215 & +0.0037 \\  
%\hline
2 & 1 & 1413 & 0.2204 $\pm$ 0.0024 & 309 & 222 & 0.1189 & +0.0011 \\  
& & (1500) & (0.2263 $\pm$ 0.0030) & (334) & (242) & (0.1204) & (+0.0026) \\  
%\hline
1/2 & 2 & 1422 & 0.2190 $\pm$ 0.0029 & 303 & 217 & 0.1185 & $+$0.0007 \\  
& & (1500) & (0.2132 $\pm$ 0.0031) & (278) & (197) & (0.1167) & 
($-$0.0011) \\  
%\hline
2 & 1/2 & 1460 & 0.2021 $\pm$ 0.0022 & 233 & 162 & 0.1134 & $-$0.0044 \\  
& & (1496) & (0.2323 $\pm$ 0.0030) & (361) & (264) & (0.1220) & (+0.0042) \\  
%\hline
1/2 & 1/2 & 1436 & 0.1975 $\pm$ 0.0012 & 216 & 149 & 0.1120 & $-$0.0058 \\  
& & (1450) & (0.1970 $\pm$ 0.0018) & (214) & (148) & (0.1120) & 
($-$0.0058) \\  
%\hline
2 & 2 & 1447 & 0.2340 $\pm$ 0.0033 & 369 & 271 & 0.1225 & +0.0047 \\  
& & (1460) & (0.2343 $\pm$ 0.0032) & (370) & (271) & (0.1226) & (+0.0048) \\  
\hline
\end{tabular}
\end{center}

\vspace{0.5cm}

We find that variations of $\asMZ$ with the variations of $k_F$ and $k_R$
are very similar to ones which have been obtained in previous subsection.
However, there is a quite big difference in the cases $k_R=2$, $k_F=1/2$ 
and $k_R=1/2$, $k_F=2$ between results in the Table 20 in
%with 
brackets and
without ones. The difference 
%that 
seems to come from the correlations between the values of 
higher-order contributions
(that is mimicked by scale dependences) and twist-four corrections, i.e.
so-called duality effect (see \cite{KPS1} and references therein).

%Moreover
As in the case of nonsinglet evolution, 
the dependence of $\asMZ$ with the variations of $k_F$ and $k_R$
is quite similar to one which have been obtained in \cite{NeVo} by the
variation of $k_R$-scale from $1/4$ to $4$.

Taking maximal and minimal values (that corresponds to 
$k_R=k_F=1/2$ and $2$, respectively) of coupling constant 
we obtain the
%following 
theoretical errors $+0.0050$ and $-0.0057$ for $\asMZ$. 
In the
case when the replacement (\ref{2.cf0}) 
has been used also in NLO corrections
to the coefficient functions (i.e. when the Eq.(\ref{2.cf0})
replaces the Eq.(\ref{2.cc0}) there) 
the theoretical uncertainties for $\asMZ$ are changed very little
but $\chi^2(F_2)$ is higher. \\

Thus, using these analyses of the SLAC, NMC, BCDMS and BFP experimental data 
for SF $F_2$ we obtain for 
\bea
\as(M_Z^2) = 0.1177 \pm 0.0007 ~\mbox{(stat)} 
\pm 0.0021 ~\mbox{(syst)}  \pm 0.0005 ~\mbox{(norm)} 
+~ \biggl\{ 
\begin{array}{l}+ 0.0047 \\ - 0.0057 \end{array} ~\mbox{(theor)},
\label{teo2}
\eea
where the theoretical uncertainties contain the scale ones (see above),
the ones due to threshold effects ($\pm 0.0001$) 
and the method error ($\pm 0.0002$) in quadratures.\\

In conclusion of the Section 
we would like to note that the theoretical uncertainties in both types
of analyses (based on nonsinglet evolution and on combined singlet and 
nonsinglet one) are
 essentially larger than the corresponding total experimental errors. \\

Indeed, 
the total experimental errors are as follows:\\
in the analyses with the nonsinglet evolution: 
\bea
& & \mbox{(total experimental error)} ~=~ \nonumber \\
& & \biggl\{ 
\begin{array}{ll} 
~\mbox{(stat)} + \mbox{(syst)} + \mbox{(norm)} 
~=~ 0.0038 & \mbox{(total linear experimental error)} \\
\sqrt{
~\mbox{(stat)}^2 + \mbox{(syst)}^2 + \mbox{(norm)}^2 } 
~=~ 0.0023 & \mbox{(total quadratic experimental error)} \end{array}
\label{exp2}
\eea
\vspace{0.5cm}
in the analyses with the combined singlet and nonsinglet evolution: 
\bea
%& & 
\mbox{(total experimental error)} ~=~ 
\biggl\{ 
\begin{array}{ll} 
%~\mbox{(stat.err.)} + \mbox{(syst.err.)} + \mbox{(norm.err.)} ~=~ 
0.0033 & \mbox{(total linear experimental error)} \\
0.0023 & \mbox{(total quadratic experimental error)}, \end{array}
\label{exp}
\eea
i.e. they are less by factor $1.5 \div 2$ to compare with corresponding
theoretical uncertainties.

%\vspace{0.5cm}

As it has been shown in \cite{NeVo,NeVo1,KPS,KPS1}, the theoretical 
uncertainties decrease
essentially (by a factor around $2.5$), when NNLO corrections have
been taken into account. So, the fits of combined data show real
necessity in analyses of DIS data at NNLO approximation.

\section{ Summary }

As a conclusion, we would like to stress again, that using the Jacobi
polynomial expansion method, developed in \cite{Barker, Kri, Kri1}, we
have studied the $Q^2$-evolution of DIS structure function $F_2$ fitting all
modern experimental data existing at 
%nonsmall 
values of Bjorken variable $x$: $x \geq 10^{-2}$. \\

{\bf 1.} From the fits we have obtained the value of the normalization 
$\asMZ$
of QCD coupling constant. First of all, we have reanalyzed the BCDMS data 
cutting the range with large systematic errors. As it is possible to see
in subsections 4.1 and 5.1 (and also the Figs. 1 and 3), 
the values of $\asMZ$ rise strongly when
the cuts of systematics were incorporated. In another side, 
the values of $\asMZ$ does not dependent on the concrete type of the
cut within 
%in the range of 
modern statistical errors.

The values $\asMZ$ obtained in various fits are
in good agreement with one other. Indeed, we have very similar
results for $\asMZ$ in separate analyses of BCDMS data (with
the cuts of systematics) and other ones. This gives us the possibility to
fit all data together.

We have found that at $Q^2 \geq 10 \div 15$ GeV$^2$ 
the formulae of pure perturbative
QCD (i.e. twist-two approximation together with target mass corrections)
are in good agreement with all data. 
The results for  $\asMZ$ are very similar for the both types of analyses:
ones, based on
nonsinglet evolution, and ones, based on combined singlet and 
nonsinglet evolution.
%ones and singlet ones. 
They have the following form:

\bea 
\zz \mbox{ from fits, based on nonsinglet evolution: } ~~~~~~ \nonumber \\
\as(M_Z^2) &=& 0.1170 \pm 0.0009 ~\mbox{(stat)}
\pm 0.0019 ~\mbox{(syst)} \pm 0.0010 ~\mbox{(norm)}, \label{re1n} \\
& & \nonumber \\
\zz \mbox{ from fits, based on combined singlet and 
nonsinglet evolution: } ~~~~~~ \nonumber \\
\as(M_Z^2) &=& 0.1180 \pm 0.0013 ~\mbox{(stat)}
\pm 0.0021 ~\mbox{(syst)} \pm 0.0009 ~\mbox{(norm)}, 
\label{re1s}
\eea

\vskip 0.5cm

When we have added twist-four corrections, we have very good agreement
between QCD (i.e. first two coefficients of Wilson expansion)
and data starting already with $Q^2 = 1$ GeV$^2$, where the Wilson
expansion should begin
%start 
to be applicable.
The results for  $\asMZ$ coincide for the both types of analyses:
%nonsinglet ones and singlet ones. They have the following form:
ones, based on
nonsinglet evolution, and ones, based on combined singlet and 
nonsinglet evolution.
%ones and singlet ones. 
They have the following form:

\bea 
\zz \mbox{ from fits, based on nonsinglet evolution: } ~~~~~~ \nonumber \\
\as(M_Z^2) &=& 0.1174 \pm 0.0007 ~\mbox{(stat)}
\pm 0.0019 ~\mbox{(syst)} \pm 0.0010 ~\mbox{(norm)}, \label{re2n} \\
& & \nonumber \\
\zz \mbox{ from fits, based on combined singlet and 
nonsinglet evolution: } ~~~~~~ \nonumber \\
\as(M_Z^2) &=& 0.1177 \pm 0.0007 ~\mbox{(stat)}
\pm 0.0021 ~\mbox{(syst)} \pm 0.0009 ~\mbox{(norm)}, 
\label{re2s}
\eea

\vskip 0.5cm

Thus, there is very good agreement (see Eqs. (\ref{re1n}), (\ref{re1s}),
(\ref{re2n}) and (\ref{re2n}))
between results based on pure perturbative QCD at quite large $Q^2$ values
(i.e. at $Q^2 \geq 10 \div 15$ GeV$^2$) and the results based on 
%fits with using of 
first two twist terms
%coefficients 
of Wilson expansion (at $Q^2 \geq 1$ GeV$^2$, 
where the Wilson expansion should  be applicable).

We would like to note that we have good agreement also with the analysis 
\cite{H1BCDMS} of
combined H1 and BCDMS data, which has been given by H1 Collaboration very 
recently. The shapes of twist-four corrections are very similar to ones
from \cite{ViMi,Liuti}.
Our results for $\as(M_Z^2)$ are in good agreement also with 
the average value for coupling constant,
%for $\asMZ$, 
presented in the recent studies (see \cite{KPS,NeVo,NeVo1,SaYnd,SSS,LEP,Bethke}
and references therein) and in
famous Bethke review \cite{Breview}.\\

{\bf 2.} As the second item of our summary we would like to note about the
real importance of NNLO corrections in analyses of DIS experimental data.
The incorporation of the NNLO corrections have been started already
several years ago in various ways (see Introduction for discussions).

The results are based on the studies of the effect of high order corrections,
%due 
which can be estimated from
the dependence of our results on factorization scale $\mu_F$ and
renormalization one $\mu_R$.
As it has been point out already in the previous Section 
 the value of the theoretical uncertainties
%error
\footnote{As it has been already shown the scale choices 
$\mu_F=\mu_R=2Q^2$ and  $\mu_F=\mu_R=Q^2/2$ give the maximal and minimal
%boundary 
values of $\as(M_Z^2)$ (at the 
various choices of values $k_F=1/2$, $k_F=2$, $k_R=1/2$ and $k_R=2$
separately) and, thus, give the basical part of theoretical error. The
additional theoretical uncertainties due to
%coming from 
our method error and 
%from
choice of threshold points are negligible.},
coming from this dependence of the results for $\as(M_Z^2)$
(given by Eqs.(\ref{teo}) and (\ref{teo2}) for two types of
$Q^2$-evolution), are equal to

\bea 
\Delta\as(M_Z^2)|_{\mbox{theo}} ~=~
\biggl\{ 
\begin{array}{l} +0.0047 \\
-0.0057 \end{array}
\label{re3}
\eea

\vskip 0.5cm

Thus, the theoretical uncertainties are
%error is
higher
essentially than the total experimental error (\ref{exp}). 
Similar values of 
the theoretical error can be found in recent analyses of DIS process
(see \cite{NeVo,NeVo1,KPS}) and of $e^+e^-$-process in \cite{LEP,Bethke}.
As it has been studied recently by van Neerven and Vogt 
%articles 
\cite{NeVo,NeVo1}, the value of theoretical error decreases strongly
%essentially
(by a factor around 2.5) when the NNLO corrections have been taken into
account. Thus, our fits of combined data performed here and also
other analyses \cite{LEP,Bethke} show real necessity to include the NNLO
corrections to the
%our 
study of DIS experimental data.

As it has been noted in Introduction, using partial information about NNLO
QCD corrections several fits of experimental data have been performed
(see \cite{PKK}-\cite{KPS1}, \cite{NeVo,NeVo1,SaYnd,AlNNLO} and references
therein).
In order to do the analyses of experimental data in full range of $x$
values, it is necessary to know exactly all NNLO QCD corrections. At present
%calculate 
three-loop corrections to 
anomalous dimensions of Wilson operators are still unknown.
These calculations, which are
known only for several finite number of fixed Mellin moments \cite{LRV},
will be performed
%done 
\cite{Verma} in nearest future by using modern approaches
(see \cite{KaKo,KaKoTMF,Verma}) to evaluate complicated Feynman 
diagrams.\\

{\bf 3.} 
At the end of our paper we would like to discuss the contributions of 
%shortly two additional subjects: the IRR-model predictions to 
higher twist corrections. 

In our study here we have reproduced well-known $x$-shape of
the twist-four corrections at the large and intermediate values of 
Bjorken variable $x$ (see, for example, the  Tables 5, 7 and 17 and
also, for example, the results of very popular article \cite{ViMi}).

We would like to note about a small-$x$ rise of the magnitude of
twist-four corrections, when we use flat parton distributions at $x \to 0$.
The rise is in full agreement with the theoretical
predictions \cite{Bartels}. As we have discussed already in the Section 5,
there is a strong correlation between the small-$x$ behavior of twist-four
corrections and the type of the corresponding asymptotics of the 
leading-twist parton distributions. The possibility to have  a
singular type of
the asymptotics leads (in our fits)
to the appearance of the rise of sea quark and gluon 
distributions as $\sim x^{-0.18}$ at low $x$ values. At this case
the rise of the magnitude
of twist-four corrections is completely
%fully 
canceled. This cancellation
is in full agreement
with theoretical and phenomenological studies and low $x$ experimental data
of H1 and L3 Collaborations (see discussions in subsection 5.3.4).\\

We would like also to give a few words concerning the IRR-model 
predictions for the twist-four and twist-six corrections.

 In our previous study \cite{KKDIS} based on the IRR-model predictions for 
%HT 
higher twist
corrections, we have found a strong correlations between these 
%HT 
corrections
and the value 
%$\as(M_Z^2)$ for normalization 
of coupling constant. The $\as(M_Z^2)$ value tends to be very small:
$\as(M_Z^2)= 0.103 \pm 0.002~\mbox{(stat)}$.
This study has been supported by fits of DELPHI
Collaboration (see \cite{DELPHI}) and by some other analyses 
%in the recent investigations
\cite{Bethke}. 
There is, however, a disagreement with the results of the paper \cite{Al1999},
where the twist-four corrections in the framework of the IRR-model do not
lead to decrease the $\as(M_Z^2)$ value.
In our opinion, the situation is not so clear here and it
needs more investigations. We hope to 
return to this problem in our future studies.

\section*{Acknowledgments}
%\vspace{-0.2cm}

Authors are grateful to Sergei Mikhailov and Alexander Nagaitsev
for useful discussions. One
of the authors (A.V.K.) was supported in part by Alexander von Humboldt
fellowship and INTAS  grant N366.

\end{document}